\documentclass[12pt,preprint]{aastex}

\slugcomment{To appear in Engle et al. 2014, \apj, 794, 80}

\shorttitle{The Secret Life of $\delta$ Cep}
\shortauthors{Engle et al.}

\begin{document}


\title{The Secret Lives of Cepheids: \\
    Evolutionary Changes and Pulsation-Induced Shock Heating in the	Prototype Classical Cepheid \object{$\delta$ Cep} \footnote{Based on observations made with the NASA/ESA \textit{Hubble Space Telescope}, obtained at the Space Telescope Science Institute, which is operated by the Association of Universities for Research in Astronomy, Inc., under NASA contract NAS 5-26555. These observations are associated with programs \#11726, \#12302 and \#13019. This work is also based on observations obtained with \textit{XMM-Newton}, an ESA science mission with instruments and contributions directly funded by ESA Member States and the USA (NASA), associated with programs \#055241, \#060374 and \#072354.}}


\author{Scott G. Engle\altaffilmark{*} and Edward F. Guinan}
\affil{Department of Astrophysics and Planetary Science, Villanova University,
    Villanova, PA 19085 USA}
\email{scott.engle@villanova.edu}

\author{Graham M. Harper}
\affil{School of Physics, Trinity College Dublin, College Green, Dublin 2 Ireland}

\author{Hilding R. Neilson}
\affil{Department of Physics \& Astronomy, East Tennessee State University, Box 70652, Johnson City, TN 37614 USA}

\and

\author{Nancy Remage Evans}
\affil{Smithsonian Astrophysical Observatory, MS 4, 60 Garden St., Cambridge, MA 02138 USA}


\altaffiltext{*}{Centre for Astronomy, James Cook University, Townsville  QLD. 4811, Australia}


\begin{abstract}
Over the past decade, the \textit{Secret Lives of Cepheids} (SLiC) program has been carried out at Villanova University to study aspects and behaviors of classical Cepheids that are still not well-understood. In this, the first of several planned papers on program Cepheids, we report the current results for \object{$\delta$ Cep}, the Cepheid prototype. Ongoing photometry has been obtained to search for changes in the pulsation period, light curve morphology and amplitude. Combining our photometry with the times of maximum light compilation by \citet{ber00} returns a small period change of dP/dt $\approx -0.1006 \pm 0.0002$ sec yr$^{\rm -1}$. There is also evidence for a gradual light amplitude increase of $\sim$0.011-mag (\textit{V}-band) and $\sim$0.012-mag (\textit{B}-band) per decade over the last $\sim50$ years. In addition, \textit{HST-COS} UV spectrophotometry and \textit{XMM-Newton} X-ray data were carried out to investigate the high-temperature plasmas present above the Cepheid photospheres. In total, from the five visits (eight exposures) with \textit{XMM-Newton}, \object{$\delta$ Cep} is found to be a soft X-ray source ($L_{\rm X}$ [0.3--2 keV] $\approx 4.5-13\times10^{\rm 28}$ erg sec$^{\rm -1}$) with peak flux at $kT$ = 0.6--0.9 keV. The X-ray activity is found to vary, possibly in phase with the stellar pulsations. From 2010--2013, nine observations of \object{$\delta$ Cep} were carried out with \textit{HST-COS}. The UV emissions are also variable, and well-phased with the stellar pulsations. Maximum UV line emissions occur near, or slightly before, maximum optical light, varying by as much as twenty times. This variability shows that pulsation-induced shock-heating plays a significant role in Cepheid atmospheres, possibly in addition to a quiescent, magnetic heating. The results of this study show Cepheid atmospheres to be rather complex and dynamic.
\end{abstract}





\section{Introduction}

Classical Cepheids (Cepheids hereafter) are arguably the most important class of variable stars.  From the \textit{Leavitt} (\textit{Period-Luminosity} $=$ \textit{P-L}) \textit{Law}, these stars serve as crucial ``standard candles'' for determining the cosmic distance scale and measuring the Hubble Constant ($H_{\rm 0}$).  Also, Cepheids play a fundamental role in the calibration of Type Ia supernovae, which indicate that the expansion of the Universe is accelerating, and also infer the existence of dark energy \citep{rie11}. In the near future, with the advent of missions such as \textit{Gaia}, \textit{JWST} and \textit{WFIRST}, Cepheids will be employed to empirically measure the Hubble constant to $\sim1\%$ \citep{fre10}.  At this precision, and combining this directly measured $H_{\rm 0}$ with the results from \textit{Planck} \citep{pla13} and \textit{WMAP} \citep{kom11} and those expected from \textit{WFIRST}, the study of Cepheids will tightly constrain fundamental cosmological parameters.

Cepheids are also important astrophysical laboratories for probing the internal structure, stellar atmospheres and evolution of moderate mass ($\sim4-15$ $M_\odot$) stars (and Type II SN progenitors, for Cepheids above $\sim8M_\odot$). For example, the study of Cepheids provides valuable information about convective core overshooting \citep{cas11}, mass loss \citep{nei12}, rotation \citep{and14} and helium core burning \citep{mor10} for moderate mass, evolved stars. Studies of changes in their pulsation periods and amplitudes reveal evolutionary changes too subtle to detect directly, and understanding these various characteristics of Cepheids is crucial to their use as high-precision standard candles.

To this end, the \textit{Secret Lives of Cepheids} (SLiC) program was initiated: a comprehensive study of Cepheid behavior, evolution, pulsations, atmospheres, heating dynamics, shocks and winds. The program started with the initially limited goal of photometrically monitoring the period and amplitude changes of Polaris \citep{dav02,nei12a}, and has since expanded to include 16 bright Cepheids, covering a wide range of pulsation properties (e.g. pulsation periods, amplitudes, modes of pulsation). The program now spans almost the entire electromagnetic spectrum, from recently acquired X-ray observations (\textit{XMM-Newton} and \textit{Chandra}), to FUV/UV spectra (\textit{HST}, \textit{IUE} and \textit{FUSE}), to the ground-based photoelectric photometry we continue to gather, to high-resolution IR spectra (\textit{SST}). Taken together, these data reveal Cepheids to be complex objects, with surprising levels of activity and behaviors. At optical wavelengths, we continue to monitor Cepheid light curves for possible signs of real-time stellar evolution, manifested by changes in light curve morphology, average brightness and times of maximum light. For example, in an increasing number of Cepheids, the pulsation period is known to change significantly, by as much as 200 sec yr$^{\rm -1}$ or more \citep{tur04,tur07}. Our photometry first indicated the recovery of Polaris' light amplitude (from its minimum value of $\sim$0.02-mag during the 1990s) in the early 2000s \citep{eng04} and subsequent studies have confirmed our initial findings \citep{spr08}. 

In addition to the evolutionary studies, we are investigating the atmospheric dynamics and heating of Cepheids using X-ray--UV data (the emphasis of this paper). \citet{fok96,gil99,gil14} (and references therein) have shown variations in the atmospheric turbulence of Cepheids due to pulsation-driven shocks, and have discussed the effects (e.g. pulsation phase-dependent photospheric line broadening). During the 1980s and 90s, studies such as \citet{sp82,sp84a,sp84b,boh94}, gathered numerous UV spectra of Cepheids with the \textit{International Ultraviolet Explorer} (\textit{IUE}), finding variable UV line emissions likely also due to shocks. \citet{sas94} further calculated that this shock heating could produce X-ray emitting plasmas. These data, combined with the current study, all show Cepheid atmospheres to contain plasmas with temperatures of 10$^{\rm 3}$ -- 10$^{\rm 7}$ K \citep{eng12,eng09} that vary in phase with the pulsations of the Cepheids, indicating that the stellar pulsations are responsible for either the very presence, or at least the excitation, of the hot plasma emissions. The most likely mechanism, as discussed later, is pulsation-driven shocks propagating through the Cepheids' outer atmospheres. The detection and characterization of shocks in Cepheid atmospheres has become rather important in recent years as they may play a role in Cepheid mass loss \citep{mar10,wil86}. The long exposures ($>$ 50 ksec) usually required for good X-ray characterizations have slowed progress in that wavelength range, but due to the excellent \textit{HST Cosmic Origins Spectrograph} (\textit{COS}) sensitivity, the UV--FUV situation is rather different. Typically, very good UV--FUV spectra are obtained in a single \textit{HST} orbit.

\section{\protect\object{$\delta$ Cep}: The Prototype of Cepheids}

\object{$\delta$ Cep} is an important calibrator for the \textit{Period-Luminosity Law} because it has one of the most precisely determined distances ($d = 273\pm 11$ pc from \textit{HST Fine Guidance Sensor} (\textit{FGS}) parallax determination) for a Cepheid \citep{ben07}. This distance agrees well with the revised \textit{Hipparcos} parallax distance of $265\pm12$-pc \citep{van07}. Selected properties of \object{$\delta$ Cep} are given in Table~\ref{tbl1}, including two separate masses derived from either stellar pulsation theory or evolution models. The difference between the two values is a prime example of the \textit{Cepheid mass discrepancy}, which has existed for decades (see \citet{nei11}). Photometry of \object{$\delta$ Cep} dates back to its discovery light curve by \citet{goo86}. It is also the second nearest Cepheid (only \object{Polaris} is nearer), and the member of a wide binary system with a hotter, A0V-type companion (\object{HD 213307}) at a projected distance of 40'' away from the Cepheid \citep{pru07}. Further, \object{$\delta$ Cep} is a member of the Cep OB6 star cluster \citep{dez99}.  \citet{maj12} determined a cluster-derived distance of $277\pm15$ pc, which agrees very well with distances derived from parallax measures. Recently, a bow shock has also been found around the Cepheid, where the stellar wind is colliding with the surrounding interstellar medium, importantly constraining the mass loss rate to 10$^{\rm -7}$--10$^{\rm -6}$ M$_\odot$ yr$^{\rm -1}$ \citep{mar10}. This was confirmed by \citet{mat12}, using \textit{Very Large Array} (\textit{VLA}) H \textsc{i} 21 cm line observations to calculate a mass loss rate of ($\sim1.0\pm0.8$) $\times10^{\rm -6}$ M$_\odot$ yr$^{\rm -1}$.

\begin{figure*}
\epsscale{0.8}
\plotone{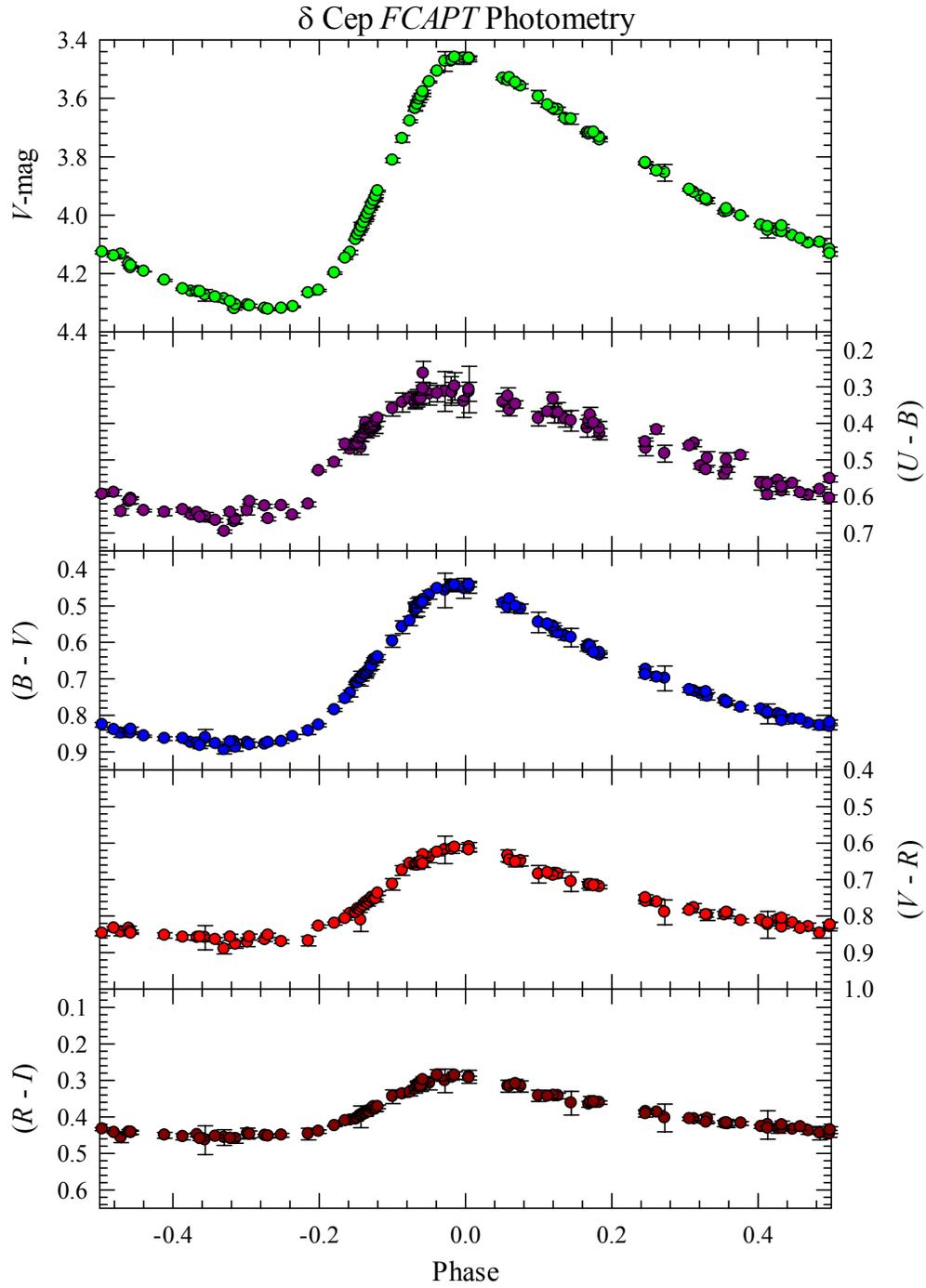}
\caption{\textit{UBVRI} photometry of \protect\object{$\delta$ Cep} obtained with the \textit{FCAPT}. The ephemeris used for phasing is that given in Table~\ref{tbl1}. \label{fig1}}
\end{figure*}

Typical of many Cepheids, \object{$\delta$ Cep} displays the ``saw tooth'' light curve, with a quicker rise to maximum brightness, and a much slower decline to minimum brightness, as shown in Fig.~\ref{fig1}. Because of its proximity (with respect to other Cepheids), \object{$\delta$ Cep} has been extensively studied at all wavelengths, including earlier UV studies (e.g. \citet{sp82,sp84a,sp84b}) carried with the \textit{International Ultraviolet Explorer} (\textit{IUE}) satellite.

\section{The Observations}


\subsection{Photometry}

\begin{figure*}
\epsscale{1}
\plotone{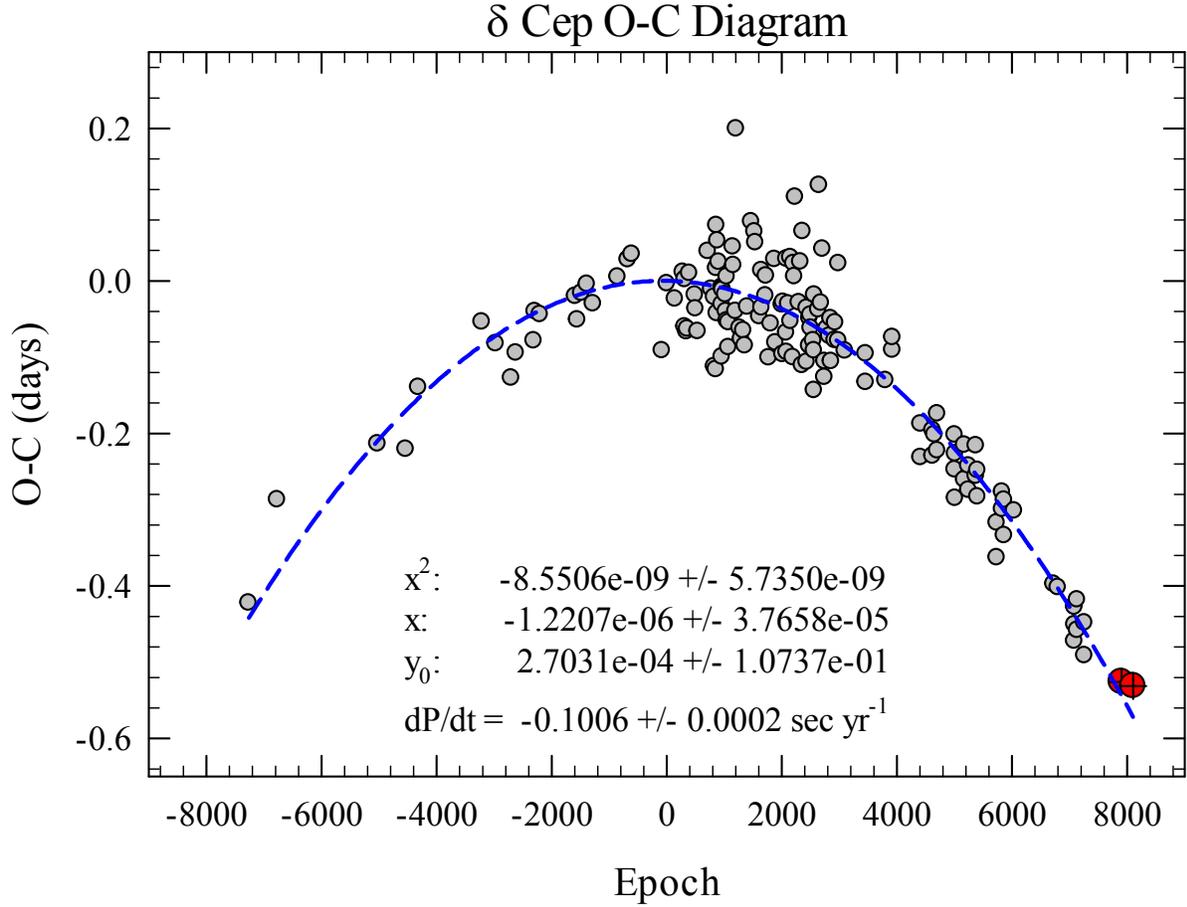}
\caption{The O-C diagram for \protect\object{$\delta$ Cep}. The region of larger scatter from epochs $\sim0-3000$ (timespan $\sim1900-1940$) represents the period of time when unaided (naked-eye) visual observations of bright variables such as \protect\object{$\delta$ Cep} were being carried out by numerous observers. The combined effects of varying observer experience, methodologies and even observing sites gives rise to the increased scatter (see \citet{ber00} for details of the archival timings dataset.) The quadratic fit parameters, which return a period change of dP/dt = $-0.1006$ sec yr$^{\rm -1}$, are given in the plot, and the steadily decreasing period trend can be seen.\label{fig2} }
\end{figure*}

Photoelectric \textit{UBVRI} photometry for this program was carried out with the \textit{Four College Automatic Photoelectric Telescope} (\textit{FCAPT} -- \citealt{ade01}) housed at \textit{Fairborn Observatory}. The full data set was split to produce two well-covered light curves, centered at epochs 2008.9 and 2011.4. Fourier analyses of the resulting light curves yielded two times of maximum light, which were added to those found in the literature \citep{ber00}. Our analysis of the complete timings dataset (Fig.~\ref{fig2}) shows the period to be decreasing over time. From a quadratic fit to the O-C data, the rate of period change is calculated to be dP/dt = $-0.1006 \pm 0.0002$ sec yr$^{\rm -1}$, indicating that \object{$\delta$ Cep} is currently making its second crossing of the instability strip \citep{tur07}. According to theory, the star contracts and heats up during the second crossing, and its surface temperature rises. Thus, the average density of the star increases, resulting in a decrease in pulsation period, according to the \textit{Period-Density Relation}
	\[P\propto\frac{1}{\sqrt{\overline{\rho}}}
\]
where \textit{P} is the pulsation period and $\overline{\rho}$ is the average stellar density. Residuals of the quadratic O-C fit show no significant, additional (cyclic or otherwise) period variability. 

\begin{figure*}
\epsscale{0.7}
\plotone{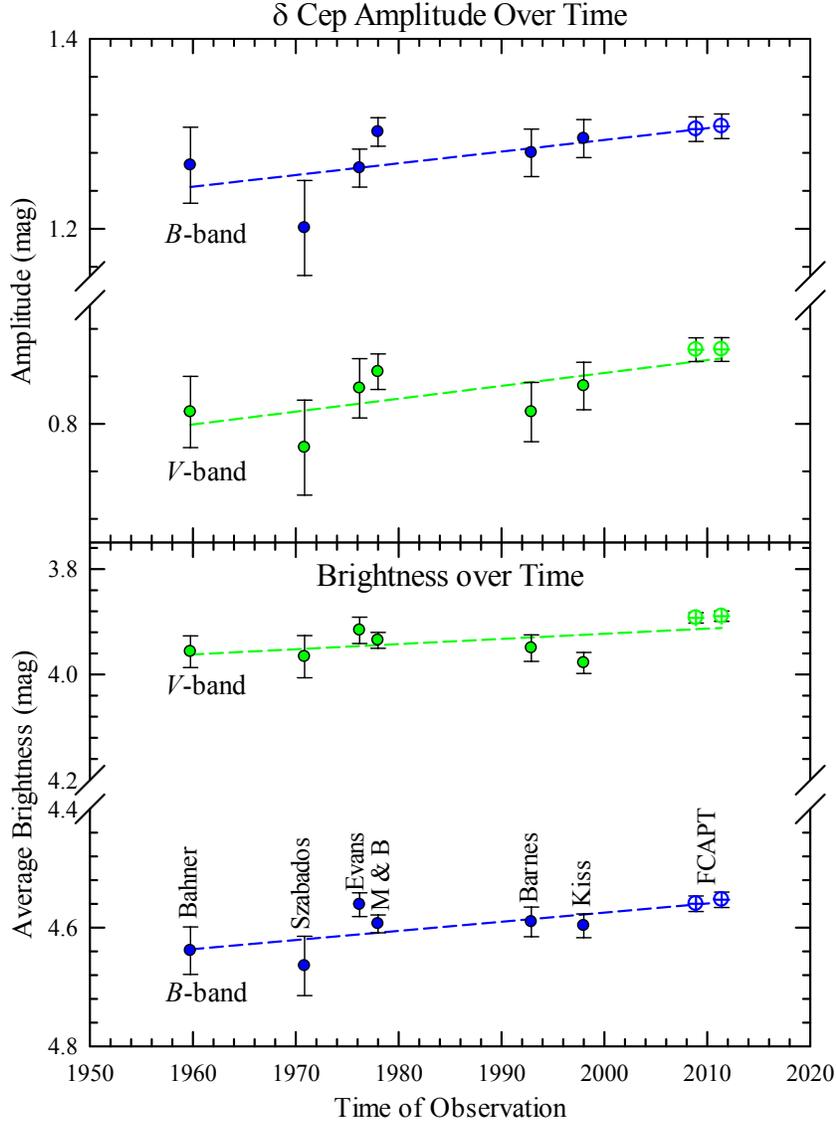}
\caption{The top panel plots light amplitudes of \protect\object{$\delta$ Cep} over time, and the bottom panel plots the average brightness of the Cepheid. Only studies where standard Johnson \textit{B}- and \textit{V}-filter data were obtained have been included in the plots. Fitting linear trends to the data show that the light amplitude of \protect\object{$\delta$ Cep} has increased at an average rates of $\sim$0.011 mag (\textit{V}) and $\sim$0.012 mag (\textit{B}) per decade. In the same time span, the average brightness of \protect\object{$\delta$ Cep} has increased by $\sim$0.01 mag (\textit{V}) and $\sim$0.02 mag (\textit{B}) per decade. Archival photometry was obtained from: \citet{kiss98,bar97,mb80,eva76,sza80,bah62}. The photometry carried out in this study made use of the same comparison stars as \citeauthor{bar97}, \citeauthor{mb80}, and \citeauthor{eva76}. \label{fig3} }
\end{figure*}

\textit{B-} and \textit{V}-band light amplitudes of $\sim1.31\pm0.013$- and $0.86\pm0.010$-mag, respectively, were also obtained from the individual \textit{FCAPT} light curves, and compared to archival observations. From this comparison, the light amplitude of \object{$\delta$ Cep} appears to be increasing over time. If simple linear trends are assumed, the light amplitude of \object{$\delta$ Cep} has been increasing at a rate of $\sim$0.011-mag in the \textit{V}-band and $\sim$0.012-mag in the \textit{B}-band, per decade, over the last $\sim50$ years (total changes of $\sim0.05$ mag in \textit{V} and $\sim0.06$ mag in \textit{B} -- Fig.~\ref{fig3}). It is interesting to note that many of the early observations of \object{$\delta$ Cep}, from the discovery of its variability in the late-1700s through to the early-1900s, display still smaller visual light amplitudes of $\sim0.5-0.7$-mag \citep{her19,mey04}. Known as a very capable and meticulous observer, \citet{ste08} observed \object{$\delta$ Cep} with a polarizing photometer and reported a visual amplitude of 0.76-mag (though a fourier analysis of the data returned an amplitude of 0.73-mag), based on observations made in 1906. These values would extend the amplitude trend observed in the past $\sim50$ years (Fig.~\ref{fig3}). However, these visual datasets require modern, careful calibrations before a conclusive determination of continued amplitude change can be achieved. In addition, the bottom panel of Fig.~\ref{fig3} shows that the average brightness of \object{$\delta$ Cep} has also increased over the same time span.

\subsection{X-ray Observations}

Constructing a theoretical model from spectroscopic observations, \citet{sas94} concluded that Cepheids could produce X-ray activity via pulsation-induced shock-heating of their atmospheres. However, pointed \textit{Einstein} and \textit{R\"{o}ntgen Satellite} (\textit{ROSAT}) observations failed to detect any such activity, and indicated that Cepheids were not (at least significant) X-ray sources. Even with \textit{log L$_{\rm X}$} $\approx$ 29 erg s$^{\rm -1}$ (as determined by \citeauthor{sas94}), the problem with detecting X-rays from Cepheids is that these stars (except for \object{Polaris} at $\sim133$ pc -- \citet{van07}) are far away, at $d > 250$ pc. Thus, they are expected to have relatively weak X-ray fluxes ($f_{\rm X} < 10^{\rm -14}$ erg s$^{\rm -1}$ cm$^{\rm -2}$). Though \object{Polaris} was detected, at the $3\sigma$ level, in a \textit{ROSAT}/\textit{High Resolution Imager} (\textit{HRI}) archival image, the detection was not noticed until several years after the observation was carried out \citep[see][]{eva07}. Confirmation of \object{Polaris}' X-ray activity and definitive detections of X-rays from other nearby Cepheids would not occur until the arrival of more powerful X-ray observatories such as \textit{XMM-Newton} and \textit{Chandra}.
	
We successfully obtained \textit{Chandra} (PI: Evans) and \textit{XMM-Newton} (PIs: Guinan \& Engle) observations of five Cepheids -- \object{Polaris}, \object{$\delta$ Cep}, \object{$\beta$ Dor}, \object{SU Cas} and \object{$\ell$~Car}  have been observed so far. The \textit{Chandra} data reduction for \object{Polaris} is discussed in \citet{eva10}. \textit{XMM-Newton} data were reduced in the usual fashion, making use of the \textit{Science Analysis System} (\textit{SAS v13.0.0}) routines. Modeling of the \textit{Chandra} and \textit{XMM-Newton} data were then carried out using the \textit{Sherpa} modeling and fitting package [distributed as part of the \textit{Chandra Interactive Analysis of Observations} (\textit{CIAO}) \textit{v4.5} suite]. \textit{MEKAL} models \citep{dra96} were used for the final two-temperature (2-T) fitting and flux calculations.

\subsection{X-ray Results}

The nearest three Cepheids (and so far the only Cepheids detected) -- \object{Polaris}, \object{$\delta$ Cep} \& \object{$\beta$ Dor} -- have X-ray luminosities of $log~L_{\rm X} \approx 28.6 - 29.2$ erg s$^{\rm -1}$. Neither the short period Cepheid \object{SU Cas} ($P = 1.95$ d; $d = 395 \pm 30$ pc) nor the long period, luminous Cepheid \object{$\ell$~Car} ($P = 35.5$ d; $d = 498 \pm 55$ pc: \citet{ben07}) were detected. Upper X-ray luminosity limits of $log~L_{\rm X} \lesssim 29.6$ and $29.5$ ergs s$^{\rm -1}$ were estimated for \object{$\ell$~Car} and \object{SU Cas}, respectively, based on exposure times, background count rates and stellar distances. Therefore, it is still possible that \object{SU Cas} and \object{$\ell$~Car} are X-ray sources with similar levels of activity to the Cepheids detected thus far, but are too distant to be detected above the background of the \textit{XMM-Newton} exposures. However, the failure to detect two of our targets underscores a long-standing ambiguity present in the X-ray studies of Cepheids. Since Cepheids are young stars ($\sim50 - 200$ Myr), any coeval main-sequence G-K-M companions (if present) would be coronal X-ray sources with X-ray luminosities similar to that of the Cepheids \citep[see][and references therein]{gui09}. Therefore, caution must be exercised as to whether unresolved companions are producing the X-ray activity. However, the \textit{HST-COS} results (\citet{eng12}, and this study) show plasmas with temperatures up to $\sim10^{\rm 5}$ K that vary (in phase) with the Cepheids' pulsation periods. This indicates that plasmas approaching soft X-ray emitting temperatures do exist in Cepheid atmospheres.

As part of this study, five separate \textit{XMM-Newton} visits were carried out for \object{$\delta$ Cep}. To provide better pulsation phase coverage, three of the visits were split during reduction into $\sim30-50$ ksec sub-exposures, resulting in 8 individual exposures for analysis. Background-subtracted count rates were $\sim0.002-0.006$ counts s$^{\rm -1}$ for all exposures. A neutral hydrogen ($N_{\rm H}$) column density value of $3.5 \times 10^{\rm 20}$ ($log~N_{\rm H} = 20.5$) cm$^{\rm -2}$ was adopted, based on distance relationships determined by \citet{par84} and corroborated by target reddening \citep{ayr05}. The relevant two-temperature \textit{MEKAL} model fitted parameters are given in Table~\ref{tbl2}. All X-ray energy distributions peak in the $kT \approx 0.6-0.9$ keV range ($\sim7-10 \times 10^{\rm 6}$ K [MK]) and have measured flux values in the range of $\sim4.5-13 \times 10^{\rm -15}$ erg s$^{\rm -1}$ cm$^{\rm -2}$ (X-ray luminosities of $log~L_{\rm X} \approx 28.6-29.1$ erg s$^{\rm -1}$). The phased X-ray fluxes are shown in Fig.~\ref{fig4}. The Cepheid's X-ray activity appears to reach a maximum near $0.5\phi$. This is different than what is found for the UV emission lines, which peak in the phase-range $0.9-1.0\phi$ (see Fig.~\ref{fig4}). The nearby A0-type companion of \object{$\delta$ Cep} -- \object{HD 213307} -- is also detected in the X-ray data, impying that this star may have a cooler, unresolved companion. For comparison, the X-ray flux of the companion star remains essentially constant in all exposures, while the X-ray flux of \object{$\delta$ Cep} increases by $2-3\times$ in the exposure near $0.5\phi$. Although our analysis has ruled out a possible flare or other transient events, this observation still represents a single (37 ksec) sub-exposure. A new (recently approved by \textit{Chandra}) observation at similar phase will further investigate the high flux level. X-ray observations of \object{$\beta$ Dor} (\textit{P} = 9.84 days), the subject of a follow-up paper, show a similar flux increase at $0.5\phi$.

Possibly related to the X-ray variations reported here for \object{$\delta$ Cep} is the recent discovery by \citet{osk14} of small, pulsation-phased X-ray variations of the $\beta$ Cep-type variable star \object{$\xi^{\rm 1}$ CMa}. The periodic X-ray variations for \object{$\xi^{\rm 1}$ CMa} (HD 46328) have the same period as the fundamental stellar pulsation of this hot, magnetic B0.5 IV star. The X-ray fluxes peak near the optical maximum brightness that occurs near the minimum stellar radius and higest temperature. The relatively strong X-ray emissions observed for $\beta$ Cep-type variables as well as other hot, massive stars arise from shock-heated winds \citep[see][]{fel97}. \citet{osk14} suggest that the periodic X-ray variations arise from small pulsation-induced changes in the wind structure, possibly coupled with changes in the magnetic field.

\subsection{Ultraviolet Spectrophotometry}

Prior to the installation of \textit{HST-COS}, the most thorough UV studies of Cepheids were carried out in the 1980s--90s with the \textit{International Ultraviolet Explorer} (\textit{IUE}). Studies of particular note are the series of papers by \citet{sp82,sp84a,sp84b} involving \textit{IUE} data ($\sim 1200-3200$\AA) for multiple Cepheids, including \object{$\delta$ Cep}, and \citet{boh94} which focused on the large available \textit{IUE} dataset for the 35 day Cepheid \object{$\ell$~Car}. These studies found UV emission lines, including O \textsc{i} 1305\AA, C \textsc{iv} 1550\AA, and He \textsc{ii} 1640\AA, to vary with the Cepheid pulsation periods -- evidence that the Cepheids (and not unseen companions) were responsible for the emissions, and pulsations may be the formation mechanism. The emissions are indicative of hot plasmas up to $100,000+$ K (also, He \textsc{ii} can be partly formed via photoionization by coronal radiation -- \citet{pag00}).

One of the major aims of this study is to improve upon previous \textit{IUE} results, since what appeared to be photospheric continuum flux introduced uncertainties in both emission line identifications and flux measures. \textit{COS}, with finer spectral resolution, higher sensitivity and lower noise, can return superior spectra for which much more precise studies could be carried out. In addition, closely-spaced emission lines that were blended together in \textit{IUE} spectra could be individually resolved in \textit{COS} data and perhaps even less-prominent emission lines would be detected. Since the only reliable data in the $\sim 1200 - 1600$\AA~ region available for Cepheids (at the beginning of this project) was low-resolution ($\sim 5-6$\AA~) \textit{IUE-SWP}(\textit{R}) spectra, it was quite frankly unknown exactly what the \textit{COS} observations would show. The \textit{HST-COS} spectra of \object{$\delta$ Cep} used in this study, and measured fluxes, are given in Table~\ref{tbl3}.

\subsection{Ultraviolet Results}

\textit{IUE} was known to suffer from scattered optical light contamination \citep{bas85}, though it turned out to be much more of an issue than was anticipated. Fig.~\ref{fig5} shows a comparison of representative \textit{IUE} and \textit{COS} spectra for \object{$\delta$ Cep}. The improvement is dramatic. The much higher resolution was expected, but as the figure shows, a great deal of what could have been continuum flux from the photosphere turned out to be scattered light. This was the reason that \textit{IUE} spectra of the program Cepheids could unambiguously show only the strongest emission lines (if any). In the case of \object{Polaris}, there was uncertainty as to whether any emission lines were present in the spectra; there was only a possible detection of the strong, but blended, oxygen/sulfur lines near $\sim 1300$\AA. The scattered light is not present in the \textit{COS} spectra of Cepheids, however, which display a wealth of emission lines not detected in the archival \textit{IUE} spectra. These lines define rich and complex Cepheid atmospheres and, as is well known, offer excellent atmospheric diagnostic potential (\citet{lin95} and references therein), since different line species originate in plasmas of specific temperatures. Also, emission line strengths and ratios, as well as line broadening and radial velocities (RVs), when measured over the stars' pulsation cycles, offer important atmospheric diagnostics and can also help distinguish Cepheid supergiant atmospheric emissions from those of possible unresolved main sequence companions (if present). The flux values reported here are integrated fluxes for the emission lines with the continuum flux subtracted. To determine RVs for the emission lines, Gaussian fits were made

Although many important emission lines are present in the \textit{COS} spectra, a few of importance were selected for the current study. These lines were selected because they represent a wide range of formation temperatures, and are relatively free from contamination via blending with nearby lines, offering a ``pure'' measurement of the emission line in question. The lines selected are: 

\begin{itemize}
\item O \textsc{i} 1358\AA~-- selected in favor of the well-known O \textsc{i} triplet at $\sim1300$\AA~ because the triplet are fluorescent lines excited by H Lyman-$\beta$ radiation \citep{kon07}, suffer from blending with (primarily) nearby S \textsc{i} lines and heavy airglow contamination. Therefore, the flux of the O triplet is not necessarily indicative of plasmas at the O \textsc{i} peak equilibrium formation temperature of $1 - 2 \times 10^{\rm 4}$ K \citep{doy97}. 
\item Si \textsc{iv} 1393/1403\AA~-- a doublet with a peak formation temperature of $\sim 5 - 8 \times 10^{\rm 4}$ K \citep{lin95}. This doublet represents an important link between the cooler, O \textsc{i} emitting plasmas and the hotter plasmas responsible for the N \textsc{v} and O \textsc{v} emission lines.
\item N \textsc{v} 1239/1243\AA~-- another doublet, but with a higher peak formation temperature of $\sim 1.5 - 2.5 \times 10^{\rm 5}$ K \citep{lin95}. This doublet is very important because the lines occur at shorter wavelengths where photospheric continuum flux is essentially negligible. The N \textsc{v} doublet is the best measure of higher-temperature plasma variability in the UV spectra of the Cepheids.
\item Additional Lines -- Although not selected for a variability study due to its location on the red wing of the Lyman-$\alpha$ geocoronal-contaminated emission line, the O \textsc{v} 1218\AA~ line is present in certain spectra and has a peak formation temperature of $\sim 2 - 4 \times 10^{\rm 5}$ K \citep{lin95}. This places it among the hottest lines observable in the UV spectra of cool stars like the Cepheids, and its detection provides an important link to higher-temperature X-ray emissions. Additionally, measures have been made of the Si \textsc{iii} 1206\AA~ and 1298\AA~ features, and the 1298/1206 flux ratio was used to carry out atmospheric density estimates.
\end{itemize}

\begin{figure*}
\epsscale{0.5}
\plotone{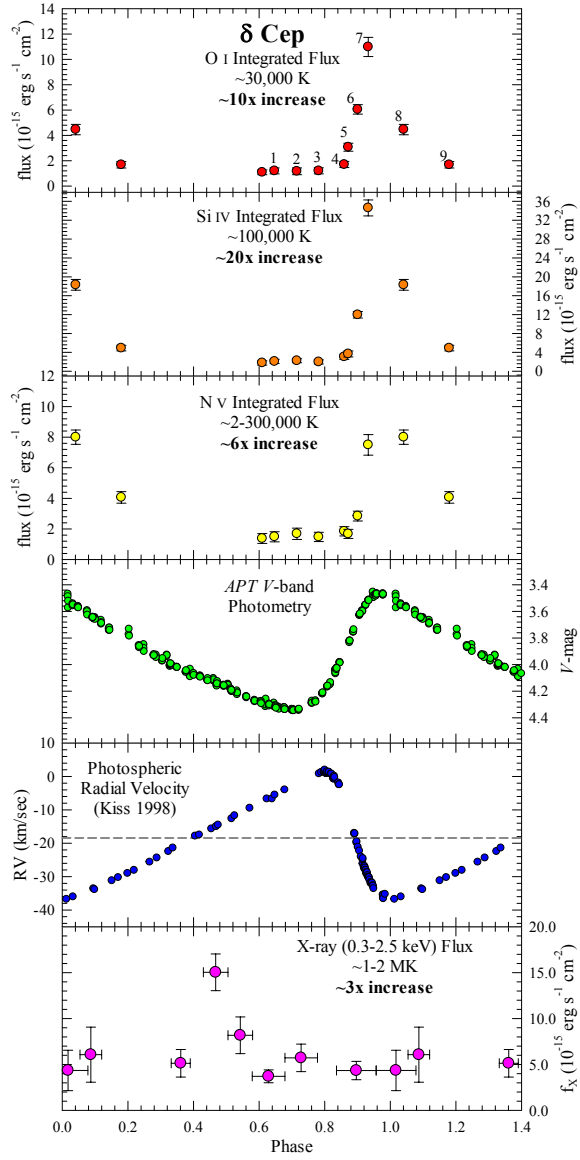}
\caption{The top three panels show integrated fluxes measured from \textit{COS} spectra of \protect\object{$\delta$ Cep}. The fourth panel shows the \textit{V}-band lightcurve obtained by us, the fifth panel shows the photospheric RVs from \citet{kiss98}, and the sixth panel shows the X-ray fluxes derived from two-temperature fits to the \textit{XMM-Newton} data. Several flux-points in the top panel are also numbered for comparison to the line profiles plotted in Fig.~\ref{fig6}.\label{fig4}}
\end{figure*}

\begin{figure*}
\epsscale{1}
\plotone{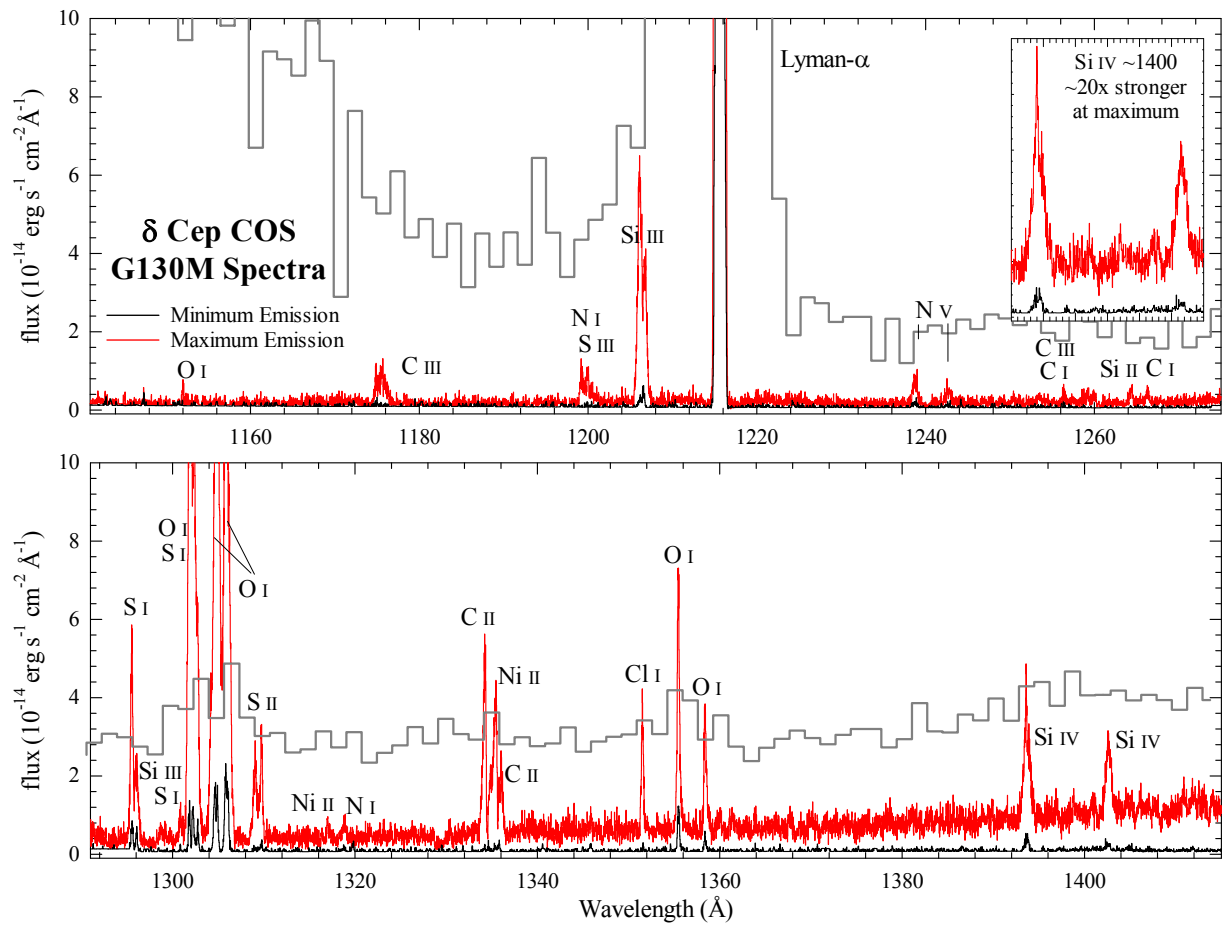}
\caption{A comparison of \textit{IUE} (gray, low-res histogram) and \textit{COS} (red and black for max/min activity levels) spectra for \protect\object{$\delta$ Cep}. Emission lines that are very well-characterized in the \textit{COS} spectra barely reveal themselves (or are absent altogether) in the scattered-light-contaminated \textit{IUE} data.\label{fig5}}
\end{figure*}

Data accumulation and results in the satellite-based high-energy studies have been much slower than for the optical studies. However, \object{$\delta$ Cep} currently has the most complete phase coverage with \textit{COS}. As an illustration of the results of the UV program thus far, the current \textit{COS} light curves of \object{$\delta$ Cep} are given in Fig.~\ref{fig4}. In looking at these plots, the most informative features are: 

\begin{enumerate}
\item The phase at which line emissions begin to increase. The phase coverage of \object{$\delta$ Cep} has been illustrative in terms of the rise in emission flux. In comparing the emission line fluxes to the included photospheric RV plot for \object{$\delta$ Cep}, we see good correlation between the phase where UV line emissions begin to peak and a ``pre-piston phase'' as given by the RVs. During this pre-piston phase, the photosphere has almost reached its minimum radius, its recession is decelerating, and it is about to begin expanding again. \citet{fok96} have shown that a shock should be propagating through the atmosphere of \object{$\delta$ Cep} at this phase ($\sim 0.8 - 0.85\phi$). The phasing of the flux increase is strong evidence in favor of a shock heating mechanism.

\item The abruptness of the increase in UV emissions from \object{$\delta$ Cep} is indicative of a sudden heating or excitation mechanism. Over a span of $\sim10$ hours (from $\sim 0.86 - 0.93\phi$), the O \textsc{i} 1358\AA~ flux increases by $\sim 7\times$ (numbered points 4 -- 7 in Fig.~\ref{fig4}) and the Si \textsc{iv} flux increases by $\sim 10\times$. Such an abrupt, strong excitation also points to a shock-related mechanism. The decrease in flux is also rapid (though not as rapid as the increase), as one would expect for a sudden heating event such as a shock.

\item The presence of an apparent quiescent level of UV emissions during $0.6 - 0.8\phi$. This indicates a persistent atmospheric heating, possibly magnetic in origin, as is the case for numerous other classes of cool stars. 

\end{enumerate}

Another feature of note is the phase-difference between the peak flux of the most energetic (highest peak formation temperature) emission feature observed -- N \textsc{v} $\sim$1240\AA~ -- and the peak fluxes of the two cooler emission features. Unfortunately, for \object{$\delta$ Cep}, this aspect of the program requires a number of \textit{very} narrowly phase-spaced observations that we simply do not have. Thus, a strict, quantitative conclusion cannot yet be drawn. However, the lower temperature plasma emission lines do appear to peak earlier than N \textsc{v}. In Fig.~\ref{fig4}, spectrum 7 is clearly the most active spectrum in terms of O \textsc{i} 1358\AA~ and Si \textsc{iv} 1393\AA~ emissions. Qualitatively speaking, spectrum 7 looks to represent the peak phase of emissions for these two features. However, for N \textsc{v} emissions, spectrum 7 has nearly the same flux as spectrum 8. Since the overall shapes of the emission curves for these three features are so similar, it can be concluded that N \textsc{v} emissions are still rising in spectrum 7, where O \textsc{i} and Si \textsc{iv} emissions have essentially peaked. Thus, N \textsc{v} must have a peak emission slightly later ($\sim0.02-0.03\phi$) in phase than O \textsc{i} and Si \textsc{iv}. As discussed by \citet{boh94}, emissions from the hottest plasmas are expected to peak first in the case of shock-heating, followed by line emissions from cooling plasmas in the post-shock regions. However, it’s worth noting that the similarity of the rise times in UV line fluxes may result from the plasma exciting line emissions of various temperatures as it begins ionizing through to higher ionization states, and the process continues as the shock front moves through more neutral and singly ionized plasma. This is an interesting behavior, but one that will require more in-depth modeling and analysis before firm conclusions are drawn.

\begin{figure*}
\epsscale{1}
\plotone{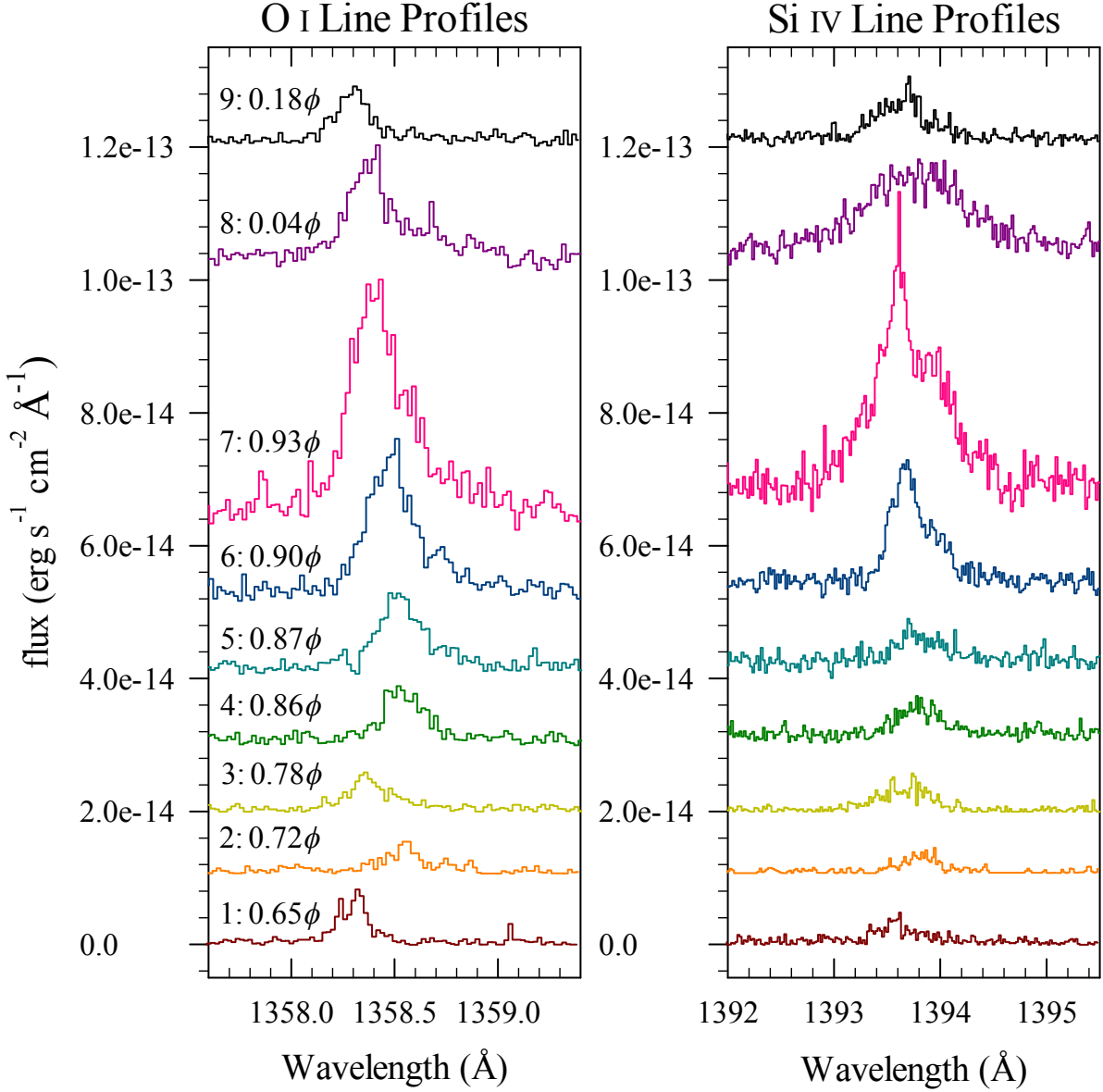}
\caption{The O \textsc{i} 1358\AA~ and Si \textsc{iv} 1393\AA~ line profiles are plotted for \protect\object{$\delta$ Cep}. The spectra are numbered, according to Fig.~\ref{fig4}. The difference in emission strengths can be seen, along with the asymmetries present at several phases, caused by the emergence of an additional blueward emission feature during phases where a shock is propagating through the atmosphere.}
\label{fig6}
\end{figure*}

In addition to what is learned from emission line fluxes about the heating mechanism(s) at work in Cepheid atmospheres, the emission line profiles and RVs can provide valuable, complementary information. Fig.~\ref{fig6} gives the profiles of the O \textsc{i} 1358\AA~ and Si \textsc{iv} 1393\AA~ emission lines observed in several spectra of \object{$\delta$ Cep}. The most potentially informative characteristic is the asymmetry present in spectra 6--8, where the lines show a strong, additional blue-shifted emission component. This component is likely caused by an expanding shock emerging from the Cepheid photosphere. On the ``near'' side of the Cepheid atmosphere, the shock is approaching, producing the blue-shifted emission. In spectrum 8, the O \textsc{i} line still shows heavy asymmetry, but the Si \textsc{iv} line shows an extremely broad and even emission profile, indicative of a large velocity distribution but no additional blue-shifted feature. At this phase ($0.04\phi$) we are likely observing Si \textsc{iv} emission from a very turbulent post-shock region, where the high turbulence is responsible for the velocity distribution, meaning that the shock has ``passed by'' the Si \textsc{iv} emitting region. The difference in O \textsc{i} and Si \textsc{iv} line profiles at this phase indicates that they are likely originating from different regions (heights) within the Cepheid atmosphere. The line profiles offer up further evidence in favor of shock-heating and compression being responsible for the enhanced emissions.

\begin{figure*}
\epsscale{0.7}
\plotone{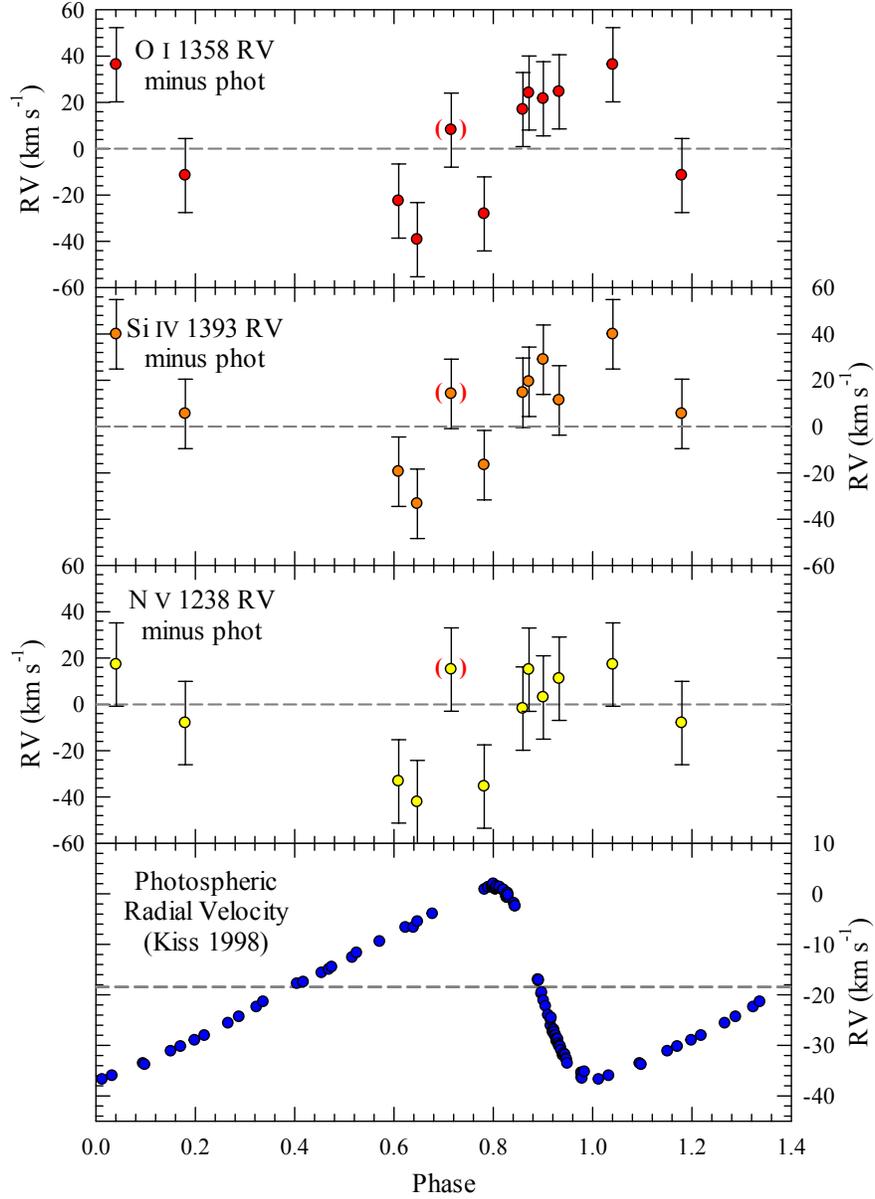}
\caption{RVs determined for the \textit{COS}-observed emission lines of \protect\object{$\delta$ Cep} are shown vs. phase. The emission line velocities have the photospheric velocities (bottom panel) removed. The bracketed RV point is from a spectrum with a possible wavelength discrepancy, but the lack of continuum flux prevents us from confirming via photospheric or ISM absorption lines. For comparison, we have again included the photospheric RVs from \citet{kiss98} in the bottom panel. \label{fig7}}
\end{figure*}

The RVs of the UV emission lines can give information on the workings of the \object{$\delta$ Cep} atmosphere, although we note there are \textit{COS} wavelength calibration issues affecting their absolute accuracy \citep{alo10}. As such, the velocities can show a larger than normal uncertainty, but the agreement in overall velocity trends between the three lines plotted gives confidence in the measures. In Fig.~\ref{fig7}, the RVs (from top to bottom panels) of the O \textsc{i}, Si \textsc{iv}, N \textsc{v} emission lines and photosphere are plotted. As indicated in the figure, the emission line RVs have had the phase-specific photospheric RV removed. For the three emission line RV plots, the dashed gray horizontal line indicates zero velocity: when the line emitting region and the photosphere have identical velocities. Thus, line RVs above this line indicate that the line emitting region is compressing on to the photosphere, and line RVs below this line indicate the region is expanding away from the photosphere. In the bottom (photospheric RV) plot, the dashed gray horizontal line represents the average velocity of the star. For spectra where the Si \textsc{iv} and/or O \textsc{i} line showed asymmetry, two Gaussian profiles were fit to the line, and the RV of the broad atmospheric emission is plotted, as opposed to the blue-shifted emission component discussed in the previous paragraph. The agreement between the Si \textsc{iv} and O \textsc{i} velocity behaviors and that of the N \textsc{v} line, which maintained a symmetric single-Gaussian profile throughout the observed phases, gives confidence in the double-Gaussian approach. 

Fig.~\ref{fig7} shows that, from $\sim0.86-1.04\phi$ (spectra 4--8 in Fig.~\ref{fig4}), the atmosphere of \object{$\delta$ Cep} is compressing on top of the photosphere. This compression begins just as the Cepheid photosphere is starting to expand again (the piston phase). This compression would result in further plasma excitation at these phases, contributing to the increased emission lines fluxes (Fig.~\ref{fig4}) and line broadening (Fig.~\ref{fig6}) that are observed.

Attempts have also been made to find suitable electron density-sensitive emission line ratios, to gain further physical insights into the Cepheid's atmosphere. A handful of well-studied ratios exist in the literature, making use of such emission lines as, e.g. C \textsc{iii} (1909\AA), Si \textsc{iii} (1892\AA) and O \textsc{iv} ($\sim$1400\AA). Unfortunately, either the spectra available for \object{$\delta$ Cep} do not cover the wavelengths of these emission lines (C \textsc{iii} 1909\AA~ or Si \textsc{iii} 1892\AA), or the lines are not strong enough to allow an unambiguous measurement (O \textsc{iv} $\sim$1400\AA). Thus, none of these density measures could successfully be applied. Making use of the latest \textsc{chianti} atomic database (http://www.chiantidatabase.org/) available at the time of writing ($v7.1.3$), investigations were carried out using a ratio of Si \textsc{iii} line fluxes (1298/1206\AA), as mentioned previously.

Measurements of almost all spectra for \object{$\delta$ Cep} give densities below the diagnostic range of the Si \textsc{iii} 1298/1206\AA~ ratio ($N_{\rm e} \lesssim6 \times 10^{\rm 9}$ cm$^{\rm -3}$). Spectrum 6 of \object{$\delta$ Cep} (as numbered in Figs.~\ref{fig4}, \ref{fig6}), where the flux is steeply rising, gives a much higher density of $N_{\rm e} \approx 3.2 \times 10^{\rm 10}$ cm$^{\rm -3}$. For reference, at similar plasma temperatures to those probed by the Si \textsc{iii} ratio above, quiet regions of the Sun have measured densities of $\sim5 \times 10^{\rm 10}$ cm$^{\rm -3}$ and active regions have densities of $\sim1 \times 10^{\rm 10}$ cm$^{\rm -3}$ \citep{dup76}, and other studies \citep[e.g.]{kee89} have also found solar densities to match that measured for spectrum 6 of \object{$\delta$ Cep}. 

However, issues have been raised with using the Si \textsc{iii} 1206\AA~ line in density diagnostics. \citet{duf83} calculated that, for the Sun, the 1206 line would be far too optically thick to give an accurate density. Although we believe that, in a supergiant such as \object{$\delta$ Cep}, the line would essentially be optically thin and suitable in that regard, there is also the temperature sensitivity to take into account when using a ground state transition, such as Si \textsc{iii} 1206\AA, with other subordinate features from higher levels. As such, we view the result as confirmation of increased atmospheric density during the phase of rising flux, but are still investigating the usefulness of the Si \textsc{iii} ratio in returning a precise, numerical density measure.

All previously-mentioned features of the spectra support the conclusion that \object{$\delta$ Cep}'s atmospheric plasmas are heated/excited via pulsation-driven compression and shock propagation.




\section{Summary \& Discussion}

The prototype Cepheid \object{$\delta$ Cep} is undergoing some very interesting and complex behaviors. Optical photometry confirms that the pulsation period of the Cepheid is steadily decreasing over time at a rate of dP/dt = $-0.1006 \pm 0.0002$ sec yr$^{\rm -1}$. The decrease in the pulsation period indicates the star's radius is slowly decreasing as it contracts and evolves toward the blue (hotter) side of the H-R diagram. Also, standardized \textit{BV} photoelectric photometry (from 1958--2012) provide evidence that the light amplitude and mean brightness are increasing. The observed changes in the light amplitude and average brightness of the star may also arise from slow evolutionary changes, indicating that we are witnessing stellar evolution in ``real time.''

Multiple X-ray observations and UV spectra have been taken of \object{$\delta$ Cep} with \textit{XMM-Newton} and \textit{HST-COS}. From these data, it is clear that \object{$\delta$ Cep} possesses an outer atmosphere of heated plasmas in the temperature range $T \approx 10^{\rm 4} - 10^{\rm 7}$ K. Soft X-ray emissions are observed ($f_{\rm X} \approx 4 - 15 \times 10^{\rm -15}$ erg s$^{\rm -1}$ cm$^{\rm -2}$) with evidence of pulsation phase-dependent variability. This variability will need to be confirmed through an approved (\textit{Chandra}), deep exposure taken at a similar phase. If the increased X-ray activity near $0.48\phi$ is confirmed through additional data, the large difference between the phases of enhanced UV and X-ray emission will imply that perhaps competing (and at times, complementary) plasma-excitation mechanisms are at play in the Cepheid atmosphere.  

The denser phasing of the \textit{HST-COS} UV spectra have provided a clearer picture of the $10^{\rm 4} - 10^{\rm 5}$ K plasmas. Though a full astrophysical modeling of the data is underway and will have to be the topic of a future UV-dedicated paper, initial results from the spectra give a clear, general picture of the \object{$\delta$ Cep} atmosphere. UV emission line fluxes vary by a factor of $\sim6-20\times$, depending on line species and plasma temperature (see Fig.~\ref{fig4}), and the rise in emission line flux begins very near the phase of minimum stellar radius, just before the photosphere begins expanding again. This is the Cepheid ``piston'' phase, when a shock is expected to emerge from the photosphere and begin propagating through the atmosphere. Shortly after this phase, as the flux is increasing, the line profiles become asymmetric due to a blue-shifted emission feature (Fig.~\ref{fig6}). This blue-shifted feature is caused by the additional excitation of plasma at the expanding shock front. During the phases of enhanced activity, the line profiles also noticeably broaden. This is a combination of the emergence of the blue-shifted feature, in addition to growing emissions from turbulent plasma in the post-shock regions and the overall compression of the atmosphere. This compression is seen in Fig.~\ref{fig7} during $\sim 0.86 - 1.04\phi$: while the photospheric recession slows, eventually turning to outward expansion, the atmosphere (as traced by emission line RVs) continues to recede, compressing around the photosphere. The phasing and strength of the flux increase, the line asymmetry and broadening of the line profiles and the photospheric vs.~atmospheric RV behaviors all confirm that the atmospheric plasmas around \object{$\delta$ Cep} undergo regular excitation from the Cepheid's radial pulsations via atmospheric compression and shock propagation. The persistence of a quiescent atmosphere, outside of the shock-excited phase-range, indicates that a second heating mechanism, such as a stellar magnetic field, is also present in \object{$\delta$ Cep}.

\section{Conclusions \& Future Perspectives}

The combined X-ray--UV--optical study of \object{$\delta$ Cep}, carried out under the SLiC program, shows the inherent complexity of this prototype Cepheid that underscores our (still incomplete) understanding of this important class of variable stars. Further, analyses of the long-term optical photometry provide strong evidence that relatively rapid evolutionary changes are taking place. This study demonstrates the high value of historical photometric data, in this case dating back to the time of \citet{goo86}.  Our study of time-series observations, from X-ray to optical wavelengths, are changing our perspective of these stars and have provided an initial step to understand not only Cepheid structures but also the details of their evolution and dynamic photospheres. We plan to continue high precision photometry to track the period change and also ascertain the observed small changes in the star's light amplitude and luminosity.

The \textit{XMM-Newton} X-ray and \textit{HST-COS} FUV spectrophotometry confirm the presence of variable, hot $\sim10^{\rm 4}$ K$- 10^{\rm 7}$ K plasmas in the outer atmosphere of this pulsating cool supergiant, most likely arising from pulsation-driven shocks. The similarity of the strengths of UV emissions (as well as X-ray luminosities) to some non-pulsating ``hybrid'' F--K giants and supergiants \citep[and references therein]{ayr11} indicate that magnetic fields could also play a role in the origin of these high-energy emission lines.  Although the presence of phase-dependent FUV line emissions was expected from prior \textit{IUE} studies of Cepheids, the X-ray emissions (especially the possible \textit{L}$_{\rm X}$ variability) is very surprising and is being followed-up.

The pulsation shock heating of the atmosphere of \object{$\delta$ Cep} (and, by extension, of other Cepheids) has yet undetermined effects on the energetics, heating, structure and dynamics of Cepheid atmospheres that could effect the \textit{P-L Law}.  As discussed previously, if Cepheids are to be used to significantly improve the cosmic distance scale and precisely determine the Hubble Constant (\textit{H}$_{\rm 0}$) to an uncertainty of $\sim$1\%, it will be necessary to have a clearer understanding of the effects of the pulsation-induced shocks on Cepheid atmospheres.  The deposition of shock-induced energies in the outer atmosphere could affect the luminosity and pulsations of the stars.  As additional targets are analyzed, the SLiC program will address these problems with a larger sample of stars.  In forthcoming papers we will present the results for additional representative Cepheids that include \object{Polaris}, \object{$\beta$ Dor}, and \object{$\ell$~Car}. Finally, we will continue to carry out ground-based photometry and to request additional space-based observations to help unlock the secret lives of these stars.



\acknowledgments

The authors wish to thank the tireless efforts of Lou Boyd, \textit{Fairborn Observatory} Director, in maintaining the \textit{FCAPT}. 

The authors also wish to thank the financial support of NSF grant AST05-07542, and NASA grants HST-GO11726, HST-GO12302, HST-GO13019 and XMM-Newton: NASA NNX14AF12G.

Support for \textit{HST} programs 11726, 12302 and 13019 was provided by NASA through grants from the Space Telescope Science Institute, which is operated by the Association of Universities for Research in Astronomy, Inc., under NASA contract NAS 5-26555.




{\it Facilities:} \facility{XMM-Newton}, \facility{Chandra}, \facility{HST (COS)}.



\begin{deluxetable}{lc}
\tablecaption{Relevant Stellar Properties of $\delta$ Cep.{Table~1\label{tbl1}}}
\tablewidth{0pt}
\tablehead{
\colhead{Parameter} & \colhead{Value}
}
\startdata
Spectral Type & F5Ib -- G1Ib$^{\rm 1}$ \\
$T_{\rm eff}$ (K) & $5500 - 6600^{\rm 1}$ \\
Mass (pulsational) ($M_\odot$) & $4.5 \pm 0.3^{\rm 2}$ \\
Mass (evolutionary) ($M_\odot$) & $5.7 \pm 0.5^{\rm 2}$ \\
Mean Luminosity ($L_\odot$) & $\sim2000^{\rm 3}$ \\
Mean Radius ($R_\odot$) & $44.5^{\rm 3}$ \\
Distance (pc) & $273 \pm 11^{\rm 4}$ \\
\textit{$\langle$V$\rangle$}-mag & $3.89\pm0.010$* \\
\textit{V}-band Amplitude & $0.86\pm0.010$* \\
\tableline
\multicolumn{2}{c}{Ephemeris (this study):}\\
\multicolumn{2}{c}{$2455479.905 + 5.366208(14) \times E$}\\
\tableline
\multicolumn{2}{c}{Ephemeris for O-C diagram \citep{ber00}:}\\
\multicolumn{2}{c}{$2412028.956 + 5.3663671 \times E$}\\
\enddata
\tablenotetext{1}{\citet{and05}}
\tablenotetext{2}{\citet{cap05}}
\tablenotetext{3}{\citet{mat12}}
\tablenotetext{4}{\citet{ben07}}
\tablenotetext{*}{this study}
\end{deluxetable}

\begin{deluxetable}{cccccccccccc}
\rotate
\tabletypesize{\scriptsize}
\setlength{\tabcolsep}{4pt}
\tablecaption{\textit{XMM-Newton} Observations of \protect\object{$\delta$ Cep}{Table~2\label{tbl2}}}
\tablewidth{0pt}
\tablehead{
\colhead{Observation} & \colhead{Start Time} & \colhead{End Time} & \colhead{Start} & \colhead{End}   & \colhead{$N_{\rm H}$}     & \colhead{Temp.} & \colhead{Temp.} & \colhead{$f_{\rm X} (0.3-2.5 keV)$} & \colhead{$f_{\rm X}$ error} & \colhead{$L_{\rm X}$} & \colhead{$log~L_{\rm X}$} \\ 

\colhead{ID \#}       & \colhead{(UT)}       & \colhead{(UT)}     & \colhead{Phase} & \colhead{Phase} & \colhead{(cm$^{\rm -2}$)} & \colhead{(keV)} & \colhead{Ratio} & \colhead{($10^{\rm -15} erg s^{\rm -1} cm^{\rm -2}$)} & \colhead{($10^{\rm -15} erg s^{\rm -1} cm^{\rm -2}$)} & \colhead{($erg s^{\rm -1}$)} & \colhead{}
}
\startdata
\dataset[ads/Sa.XMM#0603740901]{0603740901} & 2006/01/19 18:04 & 2006/01/20 12:37 & 0.054 & 0.12 & & 0.254 & 1.00 & 6.070 & 3.000 & 5.382E+28 & 28.73 \\
 & 2455217.253 & 2455218.026 &  &  &  & 0.704 & 0.63 &  &  &  &  \\
 &  &  &  &  &  &  &  &  &  &  &  \\
\dataset[ads/Sa.XMM#0603741001]{0603741001\_I} & 2006/01/21 18:05 & 2006/01/22 14:17 & 0.43 & 0.51 &  & 0.618 & 0.29 & 15.040 & 2.000 & 1.334E+29 & 29.13 \\
 & 2455219.254 & 2455220.095 &  &  &  & 2.115 & 1.00 &  &  &  &  \\
 &  &  &  &  &  &  &  &  &  &  &  \\
\dataset[ads/Sa.XMM#0603741001]{0603741001\_II} &  &  & 0.51 & 0.58 &  & 0.353 & 0.48 & 8.172 & 2.000 & 7.246E+28 & 28.86 \\
 &  &  &  &  &  & 1.345 & 1.00 &  &  &  &  \\
 &  &  &  &  &  &  &  &  &  &  &  \\
\dataset[ads/Sa.XMM#0552410401]{0552410401} & 2004/06/4 14:26 & 2004/06/4 21:53 & 0.332 & 0.39 &  & 0.995 & 1.00 & 5.130 & 1.500 & 4.549E+28 & 28.66 \\
 & 2454623.101 & 2454623.412 &  &  & 3.5E+20 & 0.401 & 0.64 &  &  &  &  \\
 &  &  &  &  &  &  &  &  &  &  &  \\
\dataset[ads/Sa.XMM#0723540301]{0723540301\_I} & 2013/06/28 6:34 & 2013/06/29 13:49 & 0.84 & 0.96 &  & 0.321 & 0.97 & 4.340 & 1.000 & 3.848E+28 & 28.59 \\
 & 2456471.774 & 2456473.076 &  &  &  & 1.328 & 1.00 &  &  &  &  \\
 &  &  &  &  &  &  &  &  &  &  &  \\
\dataset[ads/Sa.XMM#0723540301]{0723540301\_II} &  &  & 0.96 & 0.08 &  & 0.408 & 1.00 & 4.350 & 2.200 & 3.857E+28 & 28.59 \\
 &  &  &  &  &  & 1.378 & 0.65 &  &  &  &  \\
 &  &  &  &  &  &  &  &  &  &  &  \\
\dataset[ads/Sa.XMM#0723540401]{0723540401\_I} & 2013/07/2 6:17 & 2013/07/3 7:51 & 0.58 & 0.68 &  & 0.613 & 1.00 & 3.710 & 0.700 & 3.290E+28 & 28.52 \\
 & 2456475.762 & 2456476.827 &  &  &  & 0.741 & 0.01 &  &  &  &  \\
 &  &  &  &  &  &  &  &  &  &  &  \\
\dataset[ads/Sa.XMM#0723540401]{0723540401\_II} &  &  & 0.68 & 0.78 &  & 0.361 & 1.00 & 5.944 & 1.500 & 5.270E+28 & 28.72 \\
 &  &  &  &  &  & 0.956 & 0.42 &  &  &  &  \\
\enddata
\tablecomments{Temp. Ratio refers to the relative contributions of each plasma temperature to the overall 2-temperature fit. They are normalized such that the more prominent temperature has a contribution factor of 1.\\Observation \#'s ending with \_I and \_II represent the first and second halves of a split observation.}
\end{deluxetable}

\begin{deluxetable}{ccccccccccccc}
\rotate
\tabletypesize{\scriptsize}
\setlength{\tabcolsep}{4pt}
\tablecaption{HST-COS Observations of \protect\object{$\delta$ Cep}{Table~3\label{tbl3}}}
\tablewidth{0pt}
\tablehead{
\colhead{Observation} & \colhead{Start Time} & \colhead{End Time} & \colhead{Phase} & \colhead{Exp. Time} & \colhead{COS}       & \colhead{Central} & \colhead{N {\sc v} 1240\AA} & \colhead{N {\sc v} Flux} 
& \colhead{O {\sc i} 1358\AA} & \colhead{O {\sc i} Flux} & \colhead{Si {\sc iv} 1400\AA} & \colhead{Si {\sc iv} Flux} \\ 

\colhead{ID \#}       & \colhead{(UT)}       & \colhead{(UT)}     & \colhead{}      & \colhead{(sec)}     & \colhead{Grating} & \colhead{$\lambda$ (\AA)}             & \colhead{Flux (Sum)}                  & \colhead{Error} 
& \colhead{Flux}              & \colhead{Error}          & \colhead{Flux (Sum)}                & \colhead{Error}
}
\startdata
\dataset[ads/Sa.HST#LBK809010]{LBK809010} & 2010/10/19 & 2010/10/19 & 0.610 & 924.992 & G130M & 1291 & 1.379E-15 & 3.203E-16 & 1.085E-15 & 1.885E-16 & 1.765E-15 & 4.519E-16 \\
 & 00:12 & 00:30 & & & & & & & & & & \\
\dataset[ads/Sa.HST#LBK809020]{LBK809020} & 2010/10/19 & 2010/10/19 &  & 923.968 & G160M & 1589 &  &  &  &  &  &  \\
 & 01:30 & 01:48 & & & & & & & & & & \\
\dataset[ads/Sa.HST#LBK817010]{LBK817010} & 2010/12/12 & 2010/12/12 & 0.715 & 924.992 & G130M & 1291 & 1.687E-15 & 3.675E-16 & 1.163E-15 & 2.658E-16 & 2.239E-15 & 4.311E-16 \\
 & 06:12 & 06:29 & & & & & & & & & & \\
\dataset[ads/Sa.HST#LBK817020]{LBK817020} & 2010/12/12 & 2010/12/12 &  & 923.904 & G160M & 1589 &  &  &  &  &  &  \\
 & 06:34 & 06:52 & & & & & & & & & & \\
\dataset[ads/Sa.HST#LBK818010]{LBK818010} & 2010/10/30 & 2010/10/30 & 0.782 & 924.96 & G130M & 1291 & 1.482E-15 & 3.015E-16 & 1.198E-15 & 2.405E-16 & 1.979E-15 & 4.326E-16 \\
 & 16:24 & 16:41 & & & & & & & & & & \\
\dataset[ads/Sa.HST#LBK818020]{LBK818020} & 2010/10/30 & 2010/10/30 &  & 923.968 & G160M & 1589 &  &  &  &  &  &  \\
 & 16:54 & 17:11 & & & & & & & & & & \\
\dataset[ads/Sa.HST#LBK819010]{LBK819010} & 2010/10/31 & 2010/10/31 & 0.872 & 925.024 & G130M & 1291 & 1.680E-15 & 2.978E-16 & 3.068E-15 & 3.159E-16 & 3.674E-15 & 6.304E-16 \\
 & 04:04 & 04:22 & & & & & & & & & & \\
\dataset[ads/Sa.HST#LBK819020]{LBK819020} & 2010/10/31 & 2010/10/31 &  & 924.032 & G160M & 1589 &  &  &  &  &  &  \\
 & 04:26 & 04:44 & & & & & & & & & & \\
\dataset[ads/Sa.HST#LBK820010]{LBK820010} & 2010/12/13 & 2010/12/14 & 0.041 & 925.024 & G130M & 1291 & 8.000E-15 & 4.657E-16 & 4.461E-15 & 4.089E-16 & 1.831E-14 & 1.142E-15 \\
 & 23:42 & 00:00 & & & & & & & & & & \\
\dataset[ads/Sa.HST#LBK820020]{LBK820020} & 2010/12/14 & 2010/12/14 &  & 924.032 & G160M & 1589 &  &  &  &  &  &  \\
 & 00:04 & 01:13 & & & & & & & & & & \\
\dataset[ads/Sa.HST#LBK821010]{LBK821010} & 2010/10/22 & 2010/10/22 & 0.180 & 925.024 & G130M & 1291 & 4.063E-15 & 3.777E-16 & 1.681E-15 & 2.540E-16 & 4.906E-15 & 5.904E-16 \\
 & 01:41 & 01:59 & & & & & & & & & & \\
\dataset[ads/Sa.HST#LBK820020]{LBK820020} & 2010/10/22 & 2010/10/22 &  & 924.032 & G160M & 1589 &  &  &  &  &  &  \\
 & 02:56 & 03:14 & & & & & & & & & & \\
\dataset[ads/Sa.HST#LBK822010]{LBK822010} & 2010/10/31 & 2010/10/31 & 0.860 & 925.024 & G130M & 1291 & 1.848E-15 & 3.016E-16 & 1.703E-15 & 2.452E-16 & 3.063E-15 & 5.278E-16 \\
 & 02:28 & 02:46 & & & & & & & & & & \\
\dataset[ads/Sa.HST#LBK822020]{LBK822020} & 2010/10/31 & 2010/10/31 &  & 924.032 & G160M & 1589 &  &  &  &  &  &  \\
 & 02:50 & 03:08 & & & & & & & & & & \\
\dataset[ads/Sa.HST#LBK823010]{LBK823010} & 2010/10/29 & 2010/10/29 & 0.647 & 924.928 & G130M & 1291 & 1.492E-15 & 3.303E-16 & 1.198E-15 & 2.363E-16 & 2.088E-15 & 4.451E-16 \\
 & 22:59 & 23:28 & & & & & & & & & & \\
\dataset[ads/Sa.HST#LBK823020]{LBK823020} & 2010/10/29 & 2010/10/29 &  & 923.936 & G160M & 1589 &  &  &  &  &  &  \\
 & 23:32 & 23:50 & & & & & & & & & & \\
\dataset[ads/Sa.HST#LBK815010]{LBK815010} & 2011/06/13 & 2011/06/13 & 0.901 & 1152.032 & G130M & 1291 & 2.845E-15 & 3.200E-16 & 6.053E-15 & 3.722E-16 & 1.197E-14 & 6.869E-16 \\
 & 16:50 & 17:12 & & & & & & & & & & \\
\dataset[ads/Sa.HST#LBK815020]{LBK815020} & 2011/06/13 & 2011/06/13 &  & 1024.032 & G160M & 1589 &  &  &  &  &  &  \\
 & 17:17 & 17:37 & & & & & & & & & & \\
\dataset[ads/Sa.HST#LC2307010]{LC2307010} & 2013/01/18 & 2013/01/18 & 0.933 & 767.008 & G130M & 1291 & 7.496E-15 & 6.732E-16 & 1.099E-14 & 7.579E-16 & 3.460E-14 & 1.696E-15 \\
 & 18:29 & 18:49 & & & & & & & & & & \\
\dataset[ads/Sa.HST#LC2307020]{LC2307020} & 2013/01/18 & 2013/01/18 &  & 763.072 & G160M & 1589 &  &  &  &  &  &  \\
 & 18:52 & 20:12 & & & & & & & & & & \\
\enddata
\tablecomments{All fluxes and errors given in erg s$^{\rm -1}$ cm$^{\rm -2}$}
\end{deluxetable}

\clearpage

\clearpage

\clearpage

\clearpage

\clearpage

\clearpage

\clearpage

\appendix

\section{Photometry}

The \textit{UBVRI} photoelectric data for \object{$\delta$ Cep}, carried out with the \textit{FCAPT} for this study, are given in Table~\ref{tbl4}. The data were obtained between December 2007 and January 2012.

\begin{deluxetable}{lrrrlrrrlrrrlrrrlrrr}
\tablecolumns{20
\tablewidth{0pc}}
\rotate
\tabletypesize{\tiny}
\setlength{\tabcolsep}{4pt}
\tablecaption{FCAPT \textit{UBVRI} Photometry of \protect\object{$\delta$ Cep}{Table~4\label{tbl4}}}
\tablehead{\colhead{HJD} & \colhead{Phase} & \colhead{V-mag} & \colhead{error} & \colhead{HJD} & \colhead{Phase} & \colhead{U-B} & \colhead{error} & \colhead{HJD} & \colhead{Phase} & \colhead{B-V} & \colhead{error} & \colhead{HJD} & \colhead{Phase} & \colhead{V-R} & \colhead{error} & \colhead{HJD} & \colhead{Phase} & \colhead{R-I} & \colhead{error} } 
\startdata
2454437.5956 & -0.2357 & 4.3136 & 0.0026 & 2454437.5926 & -0.2362 & 0.6518 & 0.0060 & 2454437.5941 & -0.2360 & 0.8588 & 0.0062 & 2454623.9116 & 0.4846 & 0.8475 & 0.0133 & 2454623.9118 & 0.4846 & 0.4449 & 0.0182 \\
2454439.5928 & 0.1365 & 3.6690 & 0.0056 & 2454439.5941 & 0.1367 & 0.3887 & 0.0216 & 2454439.5928 & 0.1365 & 0.5839 & 0.0104 & 2454624.9074 & -0.3299 & 0.8906 & 0.0133 & 2454624.9076 & -0.3298 & 0.4568 & 0.0213 \\
2454440.5858 & 0.3215 & 3.9369 & 0.0015 & 2454440.5856 & 0.3215 & 0.5173 & 0.0081 & 2454440.5857 & 0.3215 & 0.7411 & 0.0058 & 2454625.9097 & -0.1431 & 0.8115 & 0.0306 & 2454625.9099 & -0.1431 & 0.3999 & 0.0298 \\
2454447.5794 & -0.3752 & 4.2614 & 0.0029 & 2454447.5806 & -0.3750 & 0.6512 & 0.0043 & 2454447.5793 & -0.3752 & 0.8752 & 0.0035 & 2454626.9008 & 0.0416 & 0.6316 & 0.1148 & 2454626.9010 & 0.0416 & 0.3114 & 0.1183 \\
2454449.5797 & -0.0024 & 3.4627 & 0.0206 & 2454449.5795 & -0.0025 & 0.3408 & 0.0429 & 2454449.5797 & -0.0024 & 0.4522 & 0.0270 & 2454628.8955 & 0.4133 & 0.8242 & 0.0368 & 2454628.8957 & 0.4134 & 0.4223 & 0.0388 \\
2454450.5797 & 0.1839 & 3.7435 & 0.0051 & 2454450.5795 & 0.1839 & 0.4299 & 0.0150 & 2454450.5797 & 0.1839 & 0.6358 & 0.0064 & 2454630.8917 & -0.2147 & 0.8686 & 0.0130 & 2454630.8919 & -0.2147 & 0.4471 & 0.0157 \\
2454452.6001 & -0.4396 & 4.1927 & 0.0008 & 2454452.6000 & -0.4396 & 0.6393 & 0.0033 & 2454452.6001 & -0.4396 & 0.8575 & 0.0028 & 2454631.8928 & -0.0281 & 0.6187 & 0.0376 & 2454631.8930 & -0.0281 & 0.3017 & 0.0322 \\
2454454.5902 & -0.0687 & 3.6357 & 0.0080 & 2454454.5914 & -0.0685 & 0.3344 & 0.0223 & 2454454.5901 & -0.0687 & 0.5130 & 0.0192 & 2454634.8827 & -0.4710 & 0.8436 & 0.0104 & 2454634.8829 & -0.4709 & 0.4573 & 0.0132 \\
2454458.6295 & -0.3160 & 4.3210 & 0.0038 & 2454458.6293 & -0.3160 & 0.6698 & 0.0066 & 2454458.6294 & -0.3160 & 0.8739 & 0.0075 & 2454636.8755 & -0.0996 & 0.7132 & 0.0159 & 2454636.8757 & -0.0996 & 0.3449 & 0.0187 \\
2454459.6007 & -0.1350 & 4.0087 & 0.0061 & 2454459.6005 & -0.1350 & 0.4212 & 0.0188 & 2454459.6007 & -0.1350 & 0.6829 & 0.0087 & 2454637.9467 & 0.1000 & 0.6852 & 0.0240 & 2454637.9469 & 0.1001 & 0.3425 & 0.0154 \\
2454460.6012 & 0.0514 & 3.5319 & 0.0044 & 2454460.6010 & 0.0514 & 0.3427 & 0.0241 & 2454460.6011 & 0.0514 & 0.4928 & 0.0098 & 2454638.8722 & 0.2725 & 0.7897 & 0.0339 & 2454638.8724 & 0.2725 & 0.4029 & 0.0381 \\
2454465.5856 & -0.0197 & 3.4726 & 0.0097 & 2454465.5854 & -0.0197 & 0.3114 & 0.0393 & 2454465.5855 & -0.0197 & 0.4432 & 0.0161 & 2454640.8665 & -0.3559 & 0.8592 & 0.0332 & 2454640.8667 & -0.3558 & 0.4639 & 0.0394 \\
2454466.5860 & 0.1667 & 3.7179 & 0.0030 & 2454466.5858 & 0.1667 & 0.4135 & 0.0243 & 2454466.5860 & 0.1667 & 0.6152 & 0.0090 & 2455098.6397 & -0.0492 & 0.6392 & 0.0147 & 2455098.6399 & -0.0492 & 0.3083 & 0.0179 \\
2454623.9115 & 0.4845 & 4.0934 & 0.0069 & 2454623.9125 & 0.4847 & 0.5818 & 0.0075 & 2454623.9114 & 0.4845 & 0.8290 & 0.0102 & 2455098.8062 & -0.0182 & 0.6168 & 0.0112 & 2455098.8064 & -0.0182 & 0.2922 & 0.0067 \\
2454624.9073 & -0.3299 & 4.2869 & 0.0101 & 2454624.9096 & -0.3295 & 0.6965 & 0.0045 & 2454624.9072 & -0.3299 & 0.8948 & 0.0110 & 2455099.6797 & 0.1446 & 0.7060 & 0.0267 & 2455099.6799 & 0.1446 & 0.3626 & 0.0325 \\
2454625.9096 & -0.1431 & 4.0377 & 0.0132 & 2454625.9119 & -0.1427 & 0.4678 & 0.0181 & 2454625.9095 & -0.1431 & 0.6995 & 0.0203 & 2455099.8099 & 0.1688 & 0.7150 & 0.0091 & 2455099.8101 & 0.1689 & 0.3659 & 0.0079 \\
2454626.9007 & 0.0416 & 3.5028 & 0.0736 & 2454626.9005 & 0.0415 & 0.3168 & 0.1136 & 2454626.9006 & 0.0416 & 0.4855 & 0.1044 & 2455100.6798 & 0.3309 & 0.7976 & 0.0147 & 2455100.6800 & 0.3310 & 0.4048 & 0.0113 \\
2454628.8954 & 0.4133 & 4.0524 & 0.0255 & 2454628.8952 & 0.4133 & 0.5968 & 0.0100 & 2454628.8953 & 0.4133 & 0.7961 & 0.0272 & 2455100.8050 & 0.3543 & 0.7972 & 0.0098 & 2455100.8052 & 0.3543 & 0.4187 & 0.0084 \\
2454630.8915 & -0.2147 & 4.2666 & 0.0064 & 2454630.8914 & -0.2147 & 0.6220 & 0.0067 & 2454630.8915 & -0.2147 & 0.8435 & 0.0091 & 2455101.8010 & -0.4601 & 0.8341 & 0.0063 & 2455101.8012 & -0.4601 & 0.4423 & 0.0075 \\
2454631.8927 & -0.0282 & 3.4735 & 0.0338 & 2454631.8925 & -0.0282 & 0.3129 & 0.0547 & 2454631.8926 & -0.0282 & 0.4578 & 0.0472 & 2455102.6711 & -0.2980 & 0.8716 & 0.0120 & 2455102.6713 & -0.2979 & 0.4458 & 0.0148 \\
2454634.8826 & -0.4710 & 4.1342 & 0.0059 & 2454634.8837 & -0.4708 & 0.6428 & 0.0085 & 2454634.8825 & -0.4710 & 0.8499 & 0.0103 & 2455102.7988 & -0.2742 & 0.8655 & 0.0043 & 2455102.7989 & -0.2742 & 0.4518 & 0.0037 \\
2454636.8754 & -0.0996 & 3.8125 & 0.0075 & 2454636.8752 & -0.0997 & 0.3607 & 0.0195 & 2454636.8753 & -0.0996 & 0.5974 & 0.0164 & 2455106.7865 & 0.4689 & 0.8297 & 0.0105 & 2455106.7867 & 0.4690 & 0.4382 & 0.0095 \\
2454637.9466 & 0.1000 & 3.5949 & 0.0221 & 2454637.9464 & 0.1000 & 0.3873 & 0.0198 & 2454637.9465 & 0.1000 & 0.5456 & 0.0281 & 2455131.7560 & 0.1220 & 0.6839 & 0.0093 & 2455131.7562 & 0.1221 & 0.3430 & 0.0087 \\
2454638.8721 & 0.2725 & 3.8548 & 0.0283 & 2454638.8719 & 0.2724 & 0.4832 & 0.0229 & 2454638.8720 & 0.2724 & 0.6989 & 0.0339 & 2455133.7728 & 0.4979 & 0.8295 & 0.0112 & 2455133.7730 & 0.4979 & 0.4465 & 0.0096 \\
2454640.8664 & -0.3559 & 4.2744 & 0.0198 & 2454640.8712 & -0.3550 & 0.6568 & 0.0127 & 2454640.8688 & -0.3554 & 0.8617 & 0.0228 & 2455134.7811 & -0.3142 & 0.8777 & 0.0105 & 2455134.7813 & -0.3142 & 0.4600 & 0.0121 \\
2455098.6395 & -0.0493 & 3.5439 & 0.0031 & 2455098.6394 & -0.0493 & 0.3202 & 0.0300 & 2455098.6395 & -0.0493 & 0.4703 & 0.0115 & 2455135.7812 & -0.1279 & 0.7493 & 0.0051 & 2455135.7814 & -0.1278 & 0.3802 & 0.0063 \\
2455098.8061 & -0.0182 & 3.4698 & 0.0098 & 2455098.8059 & -0.0183 & 0.3162 & 0.0282 & 2455098.8060 & -0.0182 & 0.4468 & 0.0147 & 2455136.7803 & 0.0583 & 0.6346 & 0.0162 & 2455136.7805 & 0.0584 & 0.3161 & 0.0164 \\
2455099.6796 & 0.1445 & 3.6712 & 0.0174 & 2455099.6794 & 0.1445 & 0.3933 & 0.0283 & 2455099.6795 & 0.1445 & 0.5873 & 0.0251 & 2455143.7568 & 0.3584 & 0.7916 & 0.0094 & 2455143.7570 & 0.3584 & 0.4198 & 0.0095 \\
2455099.8098 & 0.1688 & 3.7231 & 0.0083 & 2455099.8096 & 0.1688 & 0.3957 & 0.0177 & 2455099.8097 & 0.1688 & 0.6079 & 0.0111 & 2455144.7440 & -0.4576 & 0.8426 & 0.0040 & 2455144.7442 & -0.4576 & 0.4436 & 0.0044 \\
2455100.6797 & 0.3309 & 3.9503 & 0.0098 & 2455100.6795 & 0.3309 & 0.4961 & 0.0188 & 2455100.6796 & 0.3309 & 0.7481 & 0.0099 & 2455145.7525 & -0.2697 & 0.8528 & 0.0057 & 2455145.7527 & -0.2697 & 0.4529 & 0.0047 \\
2455100.8049 & 0.3542 & 3.9894 & 0.0057 & 2455100.8047 & 0.3542 & 0.5408 & 0.0110 & 2455100.8049 & 0.3543 & 0.7592 & 0.0064 & 2455146.7368 & -0.0863 & 0.6751 & 0.0133 & 2455146.7370 & -0.0862 & 0.3376 & 0.0091 \\
2455101.8009 & -0.4601 & 4.1662 & 0.0051 & 2455101.8007 & -0.4602 & 0.6152 & 0.0068 & 2455101.8008 & -0.4602 & 0.8436 & 0.0056 & 2455151.7215 & -0.1574 & 0.7938 & 0.0089 & 2455151.7217 & -0.1573 & 0.4094 & 0.0075 \\
2455102.6697 & -0.2982 & 4.3076 & 0.0037 & 2455102.6695 & -0.2983 & 0.6399 & 0.0111 & 2455102.6696 & -0.2983 & 0.8748 & 0.0098 & 2455468.8626 & -0.0577 & 0.6324 & 0.0083 & 2455468.8628 & -0.0577 & 0.3048 & 0.0098 \\
2455102.7986 & -0.2742 & 4.3196 & 0.0036 & 2455102.7984 & -0.2743 & 0.6272 & 0.0052 & 2455102.7986 & -0.2742 & 0.8799 & 0.0044 & 2455469.8525 & 0.1268 & 0.6848 & 0.0109 & 2455469.8527 & 0.1268 & 0.3427 & 0.0078 \\
2455106.7864 & 0.4689 & 4.0959 & 0.0071 & 2455106.7862 & 0.4689 & 0.5972 & 0.0116 & 2455106.7863 & 0.4689 & 0.8225 & 0.0090 & 2455470.8497 & 0.3126 & 0.7778 & 0.0119 & 2455470.8499 & 0.3126 & 0.4063 & 0.0100 \\
2455131.7559 & 0.1220 & 3.6400 & 0.0047 & 2455131.7557 & 0.1220 & 0.3726 & 0.0261 & 2455131.7558 & 0.1220 & 0.5682 & 0.0131 & 2455471.8479 & 0.4986 & 0.8250 & 0.0085 & 2455471.8481 & 0.4987 & 0.4364 & 0.0092 \\
2455133.7727 & 0.4978 & 4.1171 & 0.0079 & 2455133.7725 & 0.4978 & 0.6057 & 0.0094 & 2455133.7726 & 0.4978 & 0.8312 & 0.0087 & 2455476.8329 & 0.4276 & 0.8102 & 0.0108 & 2455476.8331 & 0.4276 & 0.4323 & 0.0127 \\
2455134.7810 & -0.3143 & 4.3066 & 0.0058 & 2455134.7808 & -0.3143 & 0.6630 & 0.0104 & 2455134.7810 & -0.3143 & 0.8879 & 0.0103 & 2455477.8318 & -0.3863 & 0.8584 & 0.0070 & 2455477.8320 & -0.3862 & 0.4547 & 0.0083 \\
2455135.7811 & -0.1279 & 3.9710 & 0.0016 & 2455135.7809 & -0.1279 & 0.4214 & 0.0131 & 2455135.7810 & -0.1279 & 0.6597 & 0.0042 & 2455478.8279 & -0.2007 & 0.8286 & 0.0038 & 2455478.8281 & -0.2006 & 0.4406 & 0.0048 \\
2455136.7802 & 0.0583 & 3.5396 & 0.0107 & 2455136.7800 & 0.0583 & 0.3263 & 0.0237 & 2455136.7801 & 0.0583 & 0.5041 & 0.0139 & 2455479.8261 & -0.0146 & 0.6120 & 0.0090 & 2455479.8263 & -0.0146 & 0.2867 & 0.0101 \\
2455143.7567 & 0.3584 & 3.9866 & 0.0033 & 2455143.7565 & 0.3583 & 0.5265 & 0.0077 & 2455143.7566 & 0.3584 & 0.7666 & 0.0068 & 2455480.8225 & 0.1710 & 0.7142 & 0.0151 & 2455480.8227 & 0.1711 & 0.3588 & 0.0147 \\
2455144.7438 & -0.4577 & 4.1803 & 0.0026 & 2455144.7449 & -0.4575 & 0.6064 & 0.0024 & 2455144.7438 & -0.4577 & 0.8490 & 0.0032 & 2455481.8207 & 0.3571 & 0.7908 & 0.0090 & 2455481.8209 & 0.3571 & 0.4168 & 0.0099 \\
2455145.7524 & -0.2697 & 4.3230 & 0.0037 & 2455145.7535 & -0.2695 & 0.6619 & 0.0052 & 2455145.7523 & -0.2697 & 0.8749 & 0.0049 & 2455482.8168 & -0.4573 & 0.8476 & 0.0022 & 2455482.8170 & -0.4573 & 0.4427 & 0.0055 \\
2455146.7367 & -0.0863 & 3.7385 & 0.0118 & 2455146.7378 & -0.0861 & 0.3430 & 0.0261 & 2455146.7366 & -0.0863 & 0.5586 & 0.0177 & 2455488.8023 & -0.3419 & 0.8648 & 0.0057 & 2455488.8025 & -0.3419 & 0.4536 & 0.0050 \\
2455151.7214 & -0.1574 & 4.1278 & 0.0068 & 2455151.7212 & -0.1574 & 0.4719 & 0.0093 & 2455151.7214 & -0.1574 & 0.7403 & 0.0095 & 2455492.8072 & 0.4044 & 0.8116 & 0.0069 & 2455492.8074 & 0.4044 & 0.4269 & 0.0051 \\
2455468.8625 & -0.0577 & 3.5880 & 0.0058 & 2455468.8623 & -0.0578 & 0.2636 & 0.0332 & 2455468.8624 & -0.0577 & 0.4836 & 0.0086 & 2455493.7962 & -0.4113 & 0.8532 & 0.0073 & 2455493.7964 & -0.4113 & 0.4506 & 0.0078 \\
2455469.8524 & 0.1267 & 3.6395 & 0.0091 & 2455469.8522 & 0.1267 & 0.3709 & 0.0277 & 2455469.8523 & 0.1267 & 0.5750 & 0.0101 & 2455495.7965 & -0.0385 & 0.6257 & 0.0089 & 2455495.7967 & -0.0385 & 0.2863 & 0.0159 \\
2455470.8496 & 0.3126 & 3.9216 & 0.0090 & 2455470.8506 & 0.3128 & 0.4550 & 0.0090 & 2455470.8495 & 0.3126 & 0.7337 & 0.0107 & 2455498.7914 & -0.4804 & 0.8334 & 0.0042 & 2455498.7916 & -0.4804 & 0.4434 & 0.0021 \\
2455471.8477 & 0.4986 & 4.1325 & 0.0074 & 2455471.8476 & 0.4986 & 0.5523 & 0.0085 & 2455471.8477 & 0.4986 & 0.8203 & 0.0075 & 2455499.7862 & -0.2950 & 0.8576 & 0.0082 & 2455499.7864 & -0.2950 & 0.4483 & 0.0096 \\
2455476.8328 & 0.4275 & 4.0543 & 0.0066 & 2455476.8326 & 0.4275 & 0.5565 & 0.0078 & 2455476.8328 & 0.4276 & 0.7966 & 0.0079 & 2455501.7760 & 0.0758 & 0.6490 & 0.0141 & 2455501.7761 & 0.0758 & 0.3168 & 0.0154 \\
2455477.8317 & -0.3863 & 4.2534 & 0.0037 & 2455477.8315 & -0.3863 & 0.6369 & 0.0067 & 2455477.8317 & -0.3863 & 0.8640 & 0.0050 & 2455502.7737 & 0.2617 & 0.7618 & 0.0091 & 2455502.7739 & 0.2617 & 0.3878 & 0.0057 \\
2455478.8278 & -0.2007 & 4.2579 & 0.0035 & 2455478.8288 & -0.2005 & 0.5306 & 0.0030 & 2455478.8277 & -0.2007 & 0.8274 & 0.0043 & 2455503.7705 & 0.4474 & 0.8195 & 0.0057 & 2455503.7707 & 0.4475 & 0.4345 & 0.0047 \\
2455479.8259 & -0.0147 & 3.4598 & 0.0058 & 2455479.8258 & -0.0147 & 0.2990 & 0.0374 & 2455479.8259 & -0.0147 & 0.4438 & 0.0124 & 2455504.7678 & -0.3667 & 0.8593 & 0.0076 & 2455504.7679 & -0.3667 & 0.4492 & 0.0054 \\
2455480.8224 & 0.1710 & 3.7180 & 0.0116 & 2455480.8222 & 0.1710 & 0.3776 & 0.0215 & 2455480.8223 & 0.1710 & 0.6115 & 0.0154 & 2455506.7644 & 0.0054 & 0.6110 & 0.0125 & 2455506.7633 & 0.0051 & 0.2903 & 0.0178 \\
2455481.8206 & 0.3570 & 3.9785 & 0.0029 & 2455481.8204 & 0.3570 & 0.5003 & 0.0195 & 2455481.8205 & 0.3570 & 0.7636 & 0.0049 & 2455508.7577 & 0.3768 & 0.8131 & 0.0051 & 2455508.7579 & 0.3768 & 0.4182 & 0.0051 \\
2455482.8167 & -0.4573 & 4.1735 & 0.0006 & 2455482.8165 & -0.4574 & 0.6105 & 0.0096 & 2455482.8166 & -0.4574 & 0.8388 & 0.0063 & 2455510.7524 & -0.2515 & 0.8705 & 0.0049 & 2455510.7526 & -0.2514 & 0.4505 & 0.0052 \\
2455488.8022 & -0.3419 & 4.2813 & 0.0050 & 2455488.8033 & -0.3417 & 0.6661 & 0.0053 & 2455488.8034 & -0.3417 & 0.8781 & 0.0051 & 2455511.7500 & -0.0656 & 0.6559 & 0.0093 & 2455511.7502 & -0.0655 & 0.3113 & 0.0094 \\
2455492.8071 & 0.4044 & 4.0341 & 0.0058 & 2455492.8069 & 0.4044 & 0.5646 & 0.0185 & 2455492.8070 & 0.4044 & 0.7843 & 0.0097 & 2455512.7464 & 0.1201 & 0.6888 & 0.0080 & 2455512.7466 & 0.1201 & 0.3422 & 0.0100 \\
2455493.7961 & -0.4113 & 4.2234 & 0.0036 & 2455493.7959 & -0.4113 & 0.6438 & 0.0093 & 2455493.7960 & -0.4113 & 0.8642 & 0.0054 & 2455513.7467 & 0.3065 & 0.7853 & 0.0058 & 2455513.7469 & 0.3065 & 0.4056 & 0.0047 \\
2455495.7963 & -0.0386 & 3.5080 & 0.0040 & 2455495.7962 & -0.0386 & 0.3188 & 0.0224 & 2455495.7963 & -0.0386 & 0.4529 & 0.0059 & 2455515.7440 & -0.3213 & 0.8583 & 0.0084 & 2455515.7442 & -0.3213 & 0.4594 & 0.0066 \\
2455498.7913 & -0.4805 & 4.1397 & 0.0038 & 2455498.7911 & -0.4805 & 0.5898 & 0.0089 & 2455498.7913 & -0.4804 & 0.8403 & 0.0063 & 2455516.7383 & -0.1360 & 0.7643 & 0.0068 & 2455516.7384 & -0.1360 & 0.3856 & 0.0085 \\
2455499.7861 & -0.2951 & 4.3109 & 0.0060 & 2455499.7872 & -0.2949 & 0.6147 & 0.0060 & 2455499.7860 & -0.2951 & 0.8810 & 0.0075 & 2455532.6063 & -0.1790 & 0.8205 & 0.0040 & 2455532.6065 & -0.1789 & 0.4245 & 0.0019 \\
2455501.7758 & 0.0757 & 3.5584 & 0.0083 & 2455502.7734 & 0.2616 & 0.4184 & 0.0107 & 2455501.7758 & 0.0757 & 0.5080 & 0.0131 & 2455532.6854 & -0.1642 & 0.8072 & 0.0039 & 2455532.6856 & -0.1642 & 0.4113 & 0.0024 \\
2455502.7736 & 0.2617 & 3.8492 & 0.0084 & 2455503.7702 & 0.4474 & 0.5650 & 0.0081 & 2455502.7735 & 0.2616 & 0.6962 & 0.0085 & 2455533.5941 & 0.0051 & 0.6192 & 0.0055 & 2455533.5943 & 0.0051 & 0.2939 & 0.0052 \\
2455503.7704 & 0.4474 & 4.0703 & 0.0041 & 2455504.7687 & -0.3666 & 0.6450 & 0.0064 & 2455503.7703 & 0.4474 & 0.8109 & 0.0073 & 2455862.6718 & 0.3292 & 0.7970 & 0.0059 & 2455862.6720 & 0.3292 & 0.4151 & 0.0039 \\
2455504.7676 & -0.3668 & 4.2629 & 0.0068 & 2455506.7628 & 0.0051 & 0.3137 & 0.0261 & 2455504.7676 & -0.3668 & 0.8764 & 0.0087 & 2455865.6113 & -0.1231 & 0.7487 & 0.0054 & 2455865.6114 & -0.1230 & 0.3780 & 0.0073 \\
2455506.7656 & 0.0056 & 3.4644 & 0.0101 & 2455508.7574 & 0.3767 & 0.4885 & 0.0094 & 2455506.7642 & 0.0053 & 0.4493 & 0.0161 & 2455866.5963 & 0.0605 & 0.6474 & 0.0064 & 2455866.5964 & 0.0605 & 0.3171 & 0.0048 \\
2455508.7576 & 0.3768 & 4.0027 & 0.0015 & 2455510.7521 & -0.2515 & 0.6259 & 0.0038 & 2455508.7575 & 0.3768 & 0.7787 & 0.0063 & 2455867.5955 & 0.2467 & 0.7607 & 0.0056 & 2455867.5957 & 0.2467 & 0.3860 & 0.0058 \\
2455510.7523 & -0.2515 & 4.3192 & 0.0031 & 2455511.7497 & -0.0656 & 0.3313 & 0.0237 & 2455510.7522 & -0.2515 & 0.8726 & 0.0042 & 2455868.5951 & 0.4330 & 0.8311 & 0.0035 & 2455868.5952 & 0.4330 & 0.4313 & 0.0053 \\
2455511.7498 & -0.0656 & 3.6148 & 0.0078 & 2455512.7461 & 0.1200 & 0.3340 & 0.0193 & 2455511.7498 & -0.0656 & 0.5041 & 0.0247 & 2455881.5652 & -0.1500 & 0.7904 & 0.0045 & 2455881.5653 & -0.1500 & 0.4069 & 0.0046 \\
2455512.7463 & 0.1201 & 3.6349 & 0.0040 & 2455513.7464 & 0.3065 & 0.4628 & 0.0073 & 2455512.7462 & 0.1201 & 0.5566 & 0.0100 & 2455881.5800 & -0.1473 & 0.7815 & 0.0030 & 2455881.5802 & -0.1472 & 0.4031 & 0.0031 \\
2455513.7465 & 0.3065 & 3.9130 & 0.0042 & 2455515.7437 & -0.3213 & 0.6444 & 0.0078 & 2455513.7465 & 0.3065 & 0.7305 & 0.0070 & 2455881.5949 & -0.1445 & 0.7834 & 0.0067 & 2455881.5950 & -0.1445 & 0.3968 & 0.0064 \\
2455515.7439 & -0.3213 & 4.2962 & 0.0067 & 2455516.7379 & -0.1361 & 0.3991 & 0.0162 & 2455515.7438 & -0.3213 & 0.8734 & 0.0084 & 2455881.6098 & -0.1417 & 0.7768 & 0.0053 & 2455881.6099 & -0.1417 & 0.3948 & 0.0038 \\
2455516.7381 & -0.1360 & 4.0154 & 0.0058 & 2455532.6060 & -0.1790 & 0.5079 & 0.0084 & 2455516.7381 & -0.1360 & 0.6895 & 0.0087 & 2455881.6246 & -0.1390 & 0.7713 & 0.0041 & 2455881.6248 & -0.1389 & 0.3922 & 0.0049 \\
2455532.6062 & -0.1790 & 4.1985 & 0.0037 & 2455532.6851 & -0.1643 & 0.4582 & 0.0140 & 2455532.6061 & -0.1790 & 0.7854 & 0.0039 & 2455881.6396 & -0.1362 & 0.7630 & 0.0063 & 2455881.6397 & -0.1361 & 0.3914 & 0.0066 \\
2455532.6853 & -0.1643 & 4.1480 & 0.0034 & 2455533.5938 & 0.0050 & 0.3076 & 0.0636 & 2455532.6852 & -0.1643 & 0.7549 & 0.0081 & 2455881.6542 & -0.1334 & 0.7657 & 0.0067 & 2455881.6543 & -0.1334 & 0.3871 & 0.0052 \\
2455533.5940 & 0.0051 & 3.4631 & 0.0044 & 2455862.6716 & 0.3291 & 0.5273 & 0.0073 & 2455533.5939 & 0.0051 & 0.4426 & 0.0056 & 2455881.6688 & -0.1307 & 0.7597 & 0.0032 & 2455881.6689 & -0.1307 & 0.3864 & 0.0047 \\
2455862.6718 & 0.3292 & 3.9454 & 0.0048 & 2455865.6121 & -0.1229 & 0.4116 & 0.0161 & 2455862.6717 & 0.3291 & 0.7361 & 0.0058 & 2455881.6834 & -0.1280 & 0.7515 & 0.0079 & 2455881.6835 & -0.1280 & 0.3776 & 0.0049 \\
2455865.6113 & -0.1231 & 3.9479 & 0.0047 & 2455866.5960 & 0.0604 & 0.3642 & 0.0151 & 2455865.6112 & -0.1231 & 0.6473 & 0.0085 & 2455881.6981 & -0.1253 & 0.7551 & 0.0035 & 2455881.6982 & -0.1252 & 0.3733 & 0.0042 \\
2455866.5962 & 0.0605 & 3.5297 & 0.0046 & 2455867.5953 & 0.2467 & 0.4690 & 0.0202 & 2455866.5962 & 0.0605 & 0.4819 & 0.0056 & 2455881.7128 & -0.1225 & 0.7531 & 0.0084 & 2455881.7129 & -0.1225 & 0.3776 & 0.0074 \\
2455867.5955 & 0.2467 & 3.8237 & 0.0048 & 2455868.5949 & 0.4329 & 0.5862 & 0.0089 & 2455867.5954 & 0.2467 & 0.6749 & 0.0078 & 2455881.7274 & -0.1198 & 0.7378 & 0.0052 & 2455881.7276 & -0.1198 & 0.3727 & 0.0050 \\
2455868.5950 & 0.4330 & 4.0575 & 0.0014 & 2455881.5649 & -0.1501 & 0.4587 & 0.0050 & 2455868.5950 & 0.4330 & 0.8013 & 0.0030 & 2455882.7387 & 0.0687 & 0.6529 & 0.0120 & 2455882.7389 & 0.0687 & 0.3096 & 0.0122 \\
2455881.5651 & -0.1500 & 4.0841 & 0.0023 & 2455881.5798 & -0.1473 & 0.4551 & 0.0141 & 2455881.5651 & -0.1500 & 0.7125 & 0.0035 & 2455888.7209 & 0.1835 & 0.7202 & 0.0049 & 2455888.7210 & 0.1835 & 0.3609 & 0.0049 \\
2455881.5800 & -0.1473 & 4.0662 & 0.0026 & 2455881.5946 & -0.1445 & 0.4413 & 0.0064 & 2455881.5799 & -0.1473 & 0.7114 & 0.0043 & 2455892.6969 & -0.0756 & 0.6567 & 0.0080 & 2455892.6971 & -0.0756 & 0.3329 & 0.0099 \\
2455881.5948 & -0.1445 & 4.0545 & 0.0033 & 2455881.6095 & -0.1418 & 0.4362 & 0.0135 & 2455881.5947 & -0.1445 & 0.7006 & 0.0052 & 2455893.7074 & 0.1127 & 0.6821 & 0.0075 & 2455893.7075 & 0.1127 & 0.3441 & 0.0047 \\
2455881.6097 & -0.1417 & 4.0402 & 0.0042 & 2455881.6244 & -0.1390 & 0.4287 & 0.0077 & 2455881.6096 & -0.1417 & 0.6987 & 0.0069 & 2455900.6875 & 0.4134 & 0.8189 & 0.0075 & 2455900.6877 & 0.4135 & 0.4316 & 0.0063 \\
2455881.6246 & -0.1390 & 4.0255 & 0.0036 & 2455881.6393 & -0.1362 & 0.4173 & 0.0093 & 2455881.6245 & -0.1390 & 0.6893 & 0.0044 & 2455911.6610 & 0.4584 & 0.8345 & 0.0057 & 2455911.6611 & 0.4584 & 0.4278 & 0.0082 \\
2455881.6395 & -0.1362 & 4.0084 & 0.0047 & 2455881.6539 & -0.1335 & 0.4129 & 0.0020 & 2455881.6394 & -0.1362 & 0.6861 & 0.0073 & 2455912.6203 & -0.3629 & 0.8583 & 0.0074 & 2455912.6204 & -0.3628 & 0.4601 & 0.0050 \\
2455881.6541 & -0.1335 & 3.9959 & 0.0056 & 2455881.6685 & -0.1308 & 0.4136 & 0.0092 & 2455881.6541 & -0.1335 & 0.6806 & 0.0058 & 2455919.5649 & -0.0687 & 0.6601 & 0.0076 & 2455919.5650 & -0.0687 & 0.3246 & 0.0104 \\
2455881.6687 & -0.1307 & 3.9808 & 0.0021 & 2455881.6832 & -0.1280 & 0.4107 & 0.0057 & 2455881.6687 & -0.1307 & 0.6726 & 0.0056 & 2455919.5806 & -0.0658 & 0.6615 & 0.0101 & 2455919.5796 & -0.0660 & 0.3160 & 0.0112 \\
2455881.6833 & -0.1280 & 3.9622 & 0.0076 & 2455881.6978 & -0.1253 & 0.4064 & 0.0085 & 2455881.6833 & -0.1280 & 0.6657 & 0.0087 & 2455919.5947 & -0.0632 & 0.6539 & 0.0117 & 2455919.5948 & -0.0632 & 0.3128 & 0.0101 \\
2455881.6980 & -0.1253 & 3.9496 & 0.0025 & 2455881.7125 & -0.1226 & 0.3926 & 0.0164 & 2455881.6980 & -0.1253 & 0.6482 & 0.0061 & 2455919.6069 & -0.0609 & 0.6510 & 0.0114 & 2455919.6070 & -0.0609 & 0.3162 & 0.0140 \\
2455881.7127 & -0.1225 & 3.9351 & 0.0048 & 2455881.7272 & -0.1198 & 0.3862 & 0.0083 & 2455881.7127 & -0.1225 & 0.6463 & 0.0061 & 2455919.6215 & -0.0582 & 0.6572 & 0.0092 & 2455919.6217 & -0.0581 & 0.2988 & 0.0087 \\
2455881.7274 & -0.1198 & 3.9171 & 0.0030 & 2455882.7385 & 0.0686 & 0.3482 & 0.0101 & 2455881.7273 & -0.1198 & 0.6407 & 0.0066 & 2455922.6360 & -0.4964 & 0.8478 & 0.0065 & 2455922.6361 & -0.4964 & 0.4338 & 0.0039 \\
2455882.7387 & 0.0687 & 3.5467 & 0.0059 & 2455888.7206 & 0.1834 & 0.4150 & 0.0201 & 2455882.7386 & 0.0686 & 0.5016 & 0.0065 & 2455926.6212 & 0.2462 & 0.7505 & 0.0086 & 2455926.6214 & 0.2463 & 0.3917 & 0.0088 \\
2455888.7209 & 0.1834 & 3.7327 & 0.0025 & 2455892.6967 & -0.0757 & 0.3329 & 0.0126 & 2455888.7208 & 0.1834 & 0.6285 & 0.0045 & 2455927.6224 & 0.4328 & 0.8066 & 0.0098 & 2455927.6225 & 0.4328 & 0.4227 & 0.0108 \\
2455892.6969 & -0.0756 & 3.6782 & 0.0051 & 2455893.7072 & 0.1127 & 0.3696 & 0.0104 & 2455892.6968 & -0.0756 & 0.5418 & 0.0128 & 2455931.6091 & 0.1757 & 0.7170 & 0.0067 & 2455931.6092 & 0.1757 & 0.3588 & 0.0071 \\
2455893.7074 & 0.1127 & 3.6233 & 0.0062 & 2455900.6873 & 0.4134 & 0.5663 & 0.0066 & 2455893.7073 & 0.1127 & 0.5509 & 0.0072 &  &  &  &  &  &  &  &  \\
2455900.6875 & 0.4134 & 4.0396 & 0.0048 & 2455911.6608 & 0.4583 & 0.5911 & 0.0063 & 2455900.6874 & 0.4134 & 0.7927 & 0.0068 &  &  &  &  &  &  &  &  \\
2455911.6610 & 0.4584 & 4.0806 & 0.0023 & 2455912.6201 & -0.3629 & 0.6577 & 0.0056 & 2455911.6609 & 0.4584 & 0.8116 & 0.0054 &  &  &  &  &  &  &  &  \\
2455912.6203 & -0.3629 & 4.2621 & 0.0061 & 2455919.5647 & -0.0688 & 0.3442 & 0.0165 & 2455912.6202 & -0.3629 & 0.8832 & 0.0077 &  &  &  &  &  &  &  &  \\
2455919.5649 & -0.0687 & 3.6346 & 0.0062 & 2455919.5793 & -0.0660 & 0.3317 & 0.0183 & 2455919.5648 & -0.0687 & 0.5022 & 0.0125 &  &  &  &  &  &  &  &  \\
2455919.5816 & -0.0656 & 3.6206 & 0.0063 & 2455919.5944 & -0.0632 & 0.3344 & 0.0118 & 2455919.5805 & -0.0658 & 0.5072 & 0.0098 &  &  &  &  &  &  &  &  \\
2455919.5946 & -0.0632 & 3.6030 & 0.0111 & 2455919.6066 & -0.0610 & 0.3326 & 0.0126 & 2455919.5946 & -0.0632 & 0.4904 & 0.0152 &  &  &  &  &  &  &  &  \\
2455919.6068 & -0.0609 & 3.5938 & 0.0103 & 2455919.6213 & -0.0582 & 0.3053 & 0.0177 & 2455919.6067 & -0.0609 & 0.4928 & 0.0126 &  &  &  &  &  &  &  &  \\
2455919.6215 & -0.0582 & 3.5778 & 0.0056 & 2455922.6358 & -0.4965 & 0.5951 & 0.0057 & 2455919.6214 & -0.0582 & 0.4911 & 0.0056 &  &  &  &  &  &  &  &  \\
2455922.6360 & -0.4964 & 4.1270 & 0.0054 & 2455926.6210 & 0.2462 & 0.4516 & 0.0113 & 2455922.6359 & -0.4964 & 0.8262 & 0.0065 &  &  &  &  &  &  &  &  \\
2455926.6211 & 0.2462 & 3.8209 & 0.0048 & 2455927.6222 & 0.4328 & 0.5737 & 0.0143 & 2455926.6211 & 0.2462 & 0.6892 & 0.0083 &  &  &  &  &  &  &  &  \\
2455927.6223 & 0.4328 & 4.0375 & 0.0060 & 2455931.6089 & 0.1757 & 0.3997 & 0.0097 & 2455927.6223 & 0.4328 & 0.8152 & 0.0090 &  &  &  &  &  &  &  &  \\
2455931.6091 & 0.1757 & 3.7166 & 0.0021 &  &  &  &  & 2455931.6090 & 0.1757 & 0.6286 & 0.0046 &  &  &  &  &  &  &  &  \\
\enddata
\end{deluxetable}

\clearpage

\end{document}